\documentclass[a4paper,10pt]{amsart}
\usepackage [centertags] {amsmath}
\usepackage {amsfonts}
\usepackage {amssymb}
\usepackage {amsmath}
\usepackage {amsthm}
\usepackage {amscd}
\usepackage [latin2] {inputenc}
\usepackage {bm, bbm}
\usepackage{xcolor}
\usepackage{cite}
\usepackage{enumerate}

\newtheorem{theorem}{Theorem}[section]
\newtheorem{lemma}[theorem]{Lemma}
\newtheorem{corollary}[theorem]{Corollary}

\newtheorem{example}[theorem]{Example}
\newtheorem{proposition}[theorem]{Proposition}

\newcommand{\hvov}{H\left( V\otimes V\right)}
\newcommand{\LVoV}{$L\left(V\otimes V\right)$ }
\newcommand{\lvov}{L\left(V\otimes V\right)}

\newcommand{\pvov}{P\left(V\otimes V\right)}

\newcommand{\pv}{P\left(V\right)}
\newcommand{\vov}{V\otimes V}
\newcommand{\fml}[1]{(\ref{fml.#1})}

\newcommand{\VoV}{$V\otimes V$ }

\newcommand{\prj}[1]{\left|#1\right>\!\left<#1\right|}
\newcommand{\pprj}[2]{\left|#1\right>\!\left<#2\right|}
\newcommand{\seq}[3]{\left\{#1\right\}_{#2}^{#3}}
\newcommand{\SEQ}[3]{$\left\{#1\right\}_{#2}^{#3} $}
\newcommand{\fnc}[2]{#1\left(#2\right)}
\newcommand{\hsqrt}[1]{\left<#1|#1\right>}
\newcommand{\hprod}[2]{\left<#1|#2\right>}

\newcommand{\id}{\mathbbm{1}}
\newcommand{\hprodA}[2]{\hprod{#2}{\fnc{#1}{#2}}}
\newcommand{\LV}{$L\left(V\right)$ }
\newcommand{\lv}{L\left(V\right)}

\newcommand{\lu}{L\left(U\right)}

\newcommand{\setC}{\mathbbm{C}}
\newcommand{\setR}{\mathbbm{R}}
\newcommand{\setN}{\mathbbm{N}}
\def\Tr{\mathop{\textnormal{Tr}}}
\newcommand{\Trb}[1]{\fnc{\Tr}{#1}}

\def\rk{\mathop{\textnormal{rk}}}

\def\mv{M\left(V\right)}

\def\muu{M\left(U\right)}

\def\mvov{M\left(V\otimes V\right)}

\newcommand{\matrfour}[1]{\left[\begin{array}{cccc}#1_{00,00} & #1_{00,01} & #1_{00,10} & #1_{00,11}\\#1_{01,00} & #1_{01,01} & #1_{01,10} & #1_{01,11}\\#1_{10,00} & #1_{10,01} & #1_{10,10} & #1_{10,11}\\#1_{11,00} & #1_{11,01} & #1_{11,10} & #1_{11,11}\end{array}\right]}

\newcommand{\convhull}[1]{\fnc{\mathop{\textnormal{convhull}}}{#1}}

\newcommand{\Repart}[1]{\fnc{\textnormal{Re}}{#1}}
\newcommand{\Impart}[1]{\fnc{\textnormal{Im}}{#1}}
\newcommand{\diag}[1]{\fnc{\textnormal{diag}}{#1}}

\newcommand{\kpositivitymaps}[1]{\mathcal{P}_{#1}}

\newcommand{\kspositivitymaps}[1]{\mathcal{SP}_{#1}}
\newcommand{\Kapositivemaps}[1]{\mathcal{P}_{#1}}
\newcommand{\Kapositivemapsb}[2]{\Kapositivemaps{#1}\left(#2\right)}
\newcommand{\PAK}[2]{P\left(#1,#2\right)}

\newcommand{\linearmapsb}[1]{L\left(#1\right)}

\newcommand{\positivitymapsb}[1]{\mathcal{P}\left(#1\right)}
\newcommand{\spositivitymapsb}[1]{\mathcal{SP}\left(#1\right)}
\newcommand{\kpositivitymapsb}[2]{\kpositivitymaps{#1}\left(#2\right)}
\newcommand{\cpositivitymapsb}[1]{\mathcal{CP}\left(#1\right)}
\newcommand{\kspositivitymapsb}[2]{\kspositivitymaps{#1}\left(#2\right)}
\newcommand{\linearopsb}[1]{L\left(#1\right)}

\newcommand{\posopsb}[1]{P\left(#1\right)}
\newcommand{\bposopsb}[1]{BP\left(#1\right)}
\newcommand{\kbposopsb}[2]{#1\textrm{-}BP\left(#2\right)}

\newcommand{\ksepopsb}[2]{#1\textrm{-}Ent\left(#2\right)}

\newcommand{\innerpr}[2]{\left<#1\right|\left.#2\right>}
\newcommand{\diad}[2]{\left|#1\right>\left<#2\right|}

\newcommand{\hconj}[1]{#1^{\ast}}
\newcommand{\hconjb}[1]{\left(#1\right)^{\ast}}

\newcommand{\identitymap}{\mathbbm{1}}
\newcommand{\identitymapn}[1]{\identitymap_{#1}}

\newcommand{\HilbertSp}{V}

\newcommand{\proj}[1]{\diad{#1}{#1}}

\newcommand{\transpos}{t}
\newcommand{\Tra}[1]{\Tr{#1}}

\title{Dualities and positivity in the study of quantum entanglement}
\author{\L ukasz Skowronek}
\address{Institute of Physics\\
Jagiellonian University\\
30-059 Krakow, Poland}
\email{lukasz.skowronek@uj.edu.pl}
\dedicatory{(a shortened and amended version of author's Master's Thesis)}

\begin{document}

\maketitle

\begin{abstract}
We present a survey on mathematical topics relating to separable states and entanglement witnesses. The convex cone duality between separable states and entanglement witnesses is discussed and later generalized to other families of operators, leading to their characterization via multiplicative properties. The condition for an operator to be an entanglement witness is rephrased as a problem of positivity of a family of real polynomials. By solving the latter in a specific case of a three-parameter family of operators, we obtain explicit description of entanglement witnesses belonging to that family. A related problem of block positivity over real numbers is discussed. We also consider a broad family of block positivity tests and prove that they can never be sufficient, which should be useful in case of future efforts in that direction. Finally, we introduce the concept of length of a separable state and present new results concerning relationships between the length and Schmidt rank. In particular, we prove that separable states of length lower of equal 3 have Schmidt ranks equal to their lengths. We also give an example of a state which has length 4 and Schmidt rank~3.

\end{abstract}

\section{Introduction}\label{sec.Introduction}
In the last two decades, a growing interest in quantum cryptography and quantum computing has significantly boosted research on entanglement in quantum systems. Despite the long lasting efforts to completely understand the mathematics of entanglement, a full characterization of $n\times m$ mixed entangled states for $nm>6$ is still missing. In particular, physicists do not know in general how to check whether a given mixed state is entangled or not. This is a renowned question in quantum information theory called the separability problem and it was the first motivation for the present work. The entanglement witness approach that we use here is not the only possible one, but it proved to be quite efficient in the past \cite{ref.Horodeccy}.

This paper is a shortened and amended version of the author's Master's Thesis\footnote{the original version is available online at {\ttfamily http://chaos.if.uj.edu.pl/\~\!karol/kzstudent.htm}  and has the title ``Quantum entanglement and certain problems in mathematics''}, defended at the Jagiellonian University in Krakow in June 2008. We give numerous insights into the separability problem in quantum mechanics, ranging from basic facts, presented in the Preliminaries section, to entirely new results, which can be found mostly in Section \ref{sec.cardinality}. A lot of the material (Sections \ref{sec.ConeDual} and \ref{sec.WitPosmaps}) has already been published elsewhere \cite{SSZ09,SZ09} in a more elaborate form, but here we take the opportunity to collect the results in a single paper, together with unpublished ones.

The paper is organized as follows. In Section \ref{sec.Preliminaries}, we give a quick outline of all the basic material needed to understand later parts of the work. We briefly discuss the mathematical structure of quantum mechanics in density matrix approach and give definitions of generalized measurements and quantum channels. Things like complete positivity, Kraus representation, separability of quantum states and the locality question in quantum mechanics have been included for the convenience of the reader. We also give an introduction about the Jamio\l kowski isomorphism, which plays an important role in later parts of the paper. 
In Section \ref{sec.ConeDual}, we start discussing the convex cone duality relating separable states to entanglement witnesses and later we generalize it to $k$-entangled states and $k$-block positive operators. Thanks to a recent work \cite{St09symmetric} by Erling St\o rmer, we know that it is possible to go even further, to general symmetric mapping cones of operators. We mention these results shortly in the end of Section \ref{sec.genposmapscr} and refer an interesed reader to St\o rmer's papers. We also make comments concerning an appealing but false conjecture about dual convex cones. In Section \ref{sec.WitPosmaps}, we show how the condition for an operator to be an entanglement witness is related to a positivity question for a family of real polynomials. We demonstrate that this correspondence can be useful by explicitly solving for entanglement witnesses belonging to a three-parameter family of operators acting on a pair of qubits. We also introduce a broad family of necessary conditions for entanglement witnesses and prove that they can never be sufficient. This is important to know because of possible efforts to obtain a sufficient condition in this way (cf. Proposition \ref{prop.insufficient} below, \cite{Sommers}). We also touch upon the subject of sums of squares and their relation to entanglement witnesses. Finally, in Section \ref{sec.cardinality}, we discuss issues related to the length of separable quantum states, which is the minimal number $l$ of products $\rho_i\otimes\sigma_i$ of positive operators  in a decomposition $\sum_{i=1}^l\rho_i\otimes\sigma_i$ of a separable state. In particular, we prove that separable quantum states of lengths $\leqslant 3$ have Schmidt ranks equal to their lengths (Proposition \ref{prop.smalllength}). By showing an example of a separable state with Schmidt rank $3$ and length $4$, we disprove a similar relation for states of bigger lengths.

Because some of the basic notation used below may differ significantly from other papers, we should mention that $L\left(V\right)$ ($H\left(V\right)$, $P\left(V\right)$, $M\left(V\right)$) denotes the set of linear operators (Hermitian operators, positive operators, quantum states, resp.) over an arbitrary (finite dimensional) linear space $V$.
Other symbols will be introduced succesively as they appear in the text.

\section{Preliminaries}\label{sec.Preliminaries}

\subsection{States \& Measurements}\label{sec.StatesMeasurements}
A paradigm of Quantum Mechanics says that one cannot in general predict results
of a single measurement on a quantum system. Nevertheless, probabilities to obtain a particular result can be predicted.  The corresponding mathematical structure is the following. All the information about a quantum system that we can extract from measurements is contained in a single entity called {\it a quantum state}. When using so-called density matrix formalism, states correspond to {\it positive, trace one} operators $\rho$ acting on a Hilbert space $\mathcal H$. Such operators $\rho$ are called {\it density matrices}. We denote the set of all of them with $M\left(\mathcal{H}\right)$.  Measurements with $n$ outcomes are represented by collections of measurement operators $\left\{A_i\right\}_{i=1}^n\subset L\left(\mathcal{H}\right)$ with the property
\begin{equation}\label{mesopcond}
 \sum_{i=1}^nA_iA_i^{\ast}=\mathbbm{1},
\end{equation}
where $\ast$ denotes the adjoint operator. More precisely, given a set of operators $\left\{A_i\right\}_{i=1}^n\subset L\left(\mathcal{H}\right)$ with the property \eqref{mesopcond} and a quantum system in a quantum state $\rho$, we can calculate the probability of obtaining the $i$-th result (say, $a_i$), 
\begin{equation}\label{probai}
 P\left(a_i\right)=\Tr\left(A_i^{\ast}\rho A_i\right).
\end{equation}
An important feature of quantum mechanics is that measurements affect the state of the system. If result $a_i$ is obtained, the system is left in the state
\begin{equation}\label{aftermeas}
\rho_i=\frac{A_i^{\ast}\rho A_i}{\Tr\left(A_i^{\ast}\rho A_i\right)}.
\end{equation}

If the result of the measurent is not recorded, but the measurement happened for sure, we may capture all the statistics of later measurements on the system by assigning to it a density matrix
\begin{equation}\label{measnonselect}
\rho_{\times}:=\sum_{i=1}^nP\left(a_i\right)\rho_i=\sum_{i=1}^nA_i^{\ast}\rho A_i.
\end{equation}
This is called a {\it nonselective measurement}.

Note that unitary transformations are included in the above formalism as a special case of a ``measurement'' with a single outcome, i.e. by taking $\left\{A_i\right\}_{i=1}^n=\left\{U\right\}$ for a unitary $U$. There is no difference between ``selective'' and ``nonselective'' unitary transformations.

A special role among quantum states is played by those represented by one-dimensional projections $\prj{\psi}$ on vectors $\psi\in\mathcal{H}$. We call them {\it pure quantum states}. They are pure in the sense that all the unpredictability  specific to them is believed to be of fundamental (quantum) nature. Similarly, measurements represented by $\left\{\prj{\psi_i}\right\}_{i=1}^d$, where $\left\{\left|\psi_i\right>\right\}_{i=1}^d$ is an orthonormal basis, should be perceived as the most fundamental ones. They are called {\it projective measurements}. After such measurement yields the $i$-th possible result, the system is left in the state $\prj{\psi_i}$. Usually, projective measurements are described as corresponding to Hermitian operators $\sum_{i=1}^na_i\prj{\psi_i}$, but this is mostly sort of a useful convention.

In everything that follows we shall assume that $\mathcal H$ is finite dimensional and thus equivalent to $\mathbbm{C}^d$ for some $d\in\mathbbm{N}$. In this case, the name of a ``Hilbert space'' is a little too exuberant. We shall make this clear in our notation by using $V$ (or $U$) instead of $\mathcal{H}$. 

\subsection{Entanglement}\label{sec.Entanglement}
In classical mechanics, the description of multi-component systems is simple at the level of constructing mathematical formalism. If the composite system consists of parts $\left\{X_i\right\}_{i=1}^n$ with degrees of freedom $x_i^j$ ($i=1,\ldots,n$, $j=1,\ldots,n_i$), the configuration space of the composite system is the Cartesian product of the configuration spaces corresponding to the individual subsystems. That is, if we need $n_i$ numbers $x_i^j$ to describe the state of the subsystem $x_i$,  it is sufficient to know all these numbers for $i=1,\ldots,n$, $j=1,\ldots,n_i$ to describe the state of the system consisting of $X_i$ with $i=1,\ldots,n$.

 In quantum mechanics, this is not the case. If the states of $X_i$ correspond to density matrices $\rho_i$ over a Hilbert space $\mathcal{H}_i$, states of the composite system consisting of $X_i$ for $i=1,\ldots,n$ are described by density matrices on the tensor product $\mathcal{H}_1\otimes\ldots\otimes\mathcal{H}_n$. 
Let $\rho$ be such a matrix, corresponding to the state of the composite system. Let $\Tr_j$ denote the trace operation in $L\left(\mathcal{H}_j\right)$. Then the {\it reduced density matrix}
\begin{equation}\label{partialtraces}
 \rho^{\left(i\right)}=\Tr_1\ldots\Tr_{i-1}\Tr_{i+1}\ldots\Tr_n\rho
\end{equation}
describes the state of the subsystem corresponding to $\mathcal{H}_i$. Since not every element of $M\left(\mathcal{H}_1\otimes\ldots\otimes\mathcal{H}_n\right)$ is of the product form $\rho^{\left(1\right)}\otimes\ldots\otimes\rho^{\left(n\right)}$, the knowledge of all the reduced matrices $\rho^{\left(i\right)}$ is not sufficient for determination of $\rho$. In other words, unlike in classical mechanics, a complete knowledge about every subsystem of a composite quantum system does not imply a complete knowledge about the composite system itself. The existence of states that are not of the form $\rho^{\left(1\right)}\otimes\ldots\otimes\rho^{\left(n\right)}$ leads to some rather non-intuitive phenomena called {\it entanglement}. Remembering that a measurement of a quantum system usually changes the state of the system (cf. formula \ref{aftermeas}), it seems that a local measurement on part $X_j$ can have an immediate influence on $X_k$ ($k\neq j$), no matter how far $X_k$ is from $X_j$. Measurements on the subsystem $X_j$ correspond to measurement operators of the form $\tilde A^{\left(j\right)}_i:=\mathbbm{1}^{\left(1\right)}\otimes\ldots\otimes A^{\left(j\right)}_i\otimes\ldots\otimes\mathbbm{1}^{\left(n\right)}$, where $A^{\left(j\right)}_i\in L\left(\mathcal{H}_j\right)$ and $\mathbbm{1}^{\left(i\right)}$ denotes the identity operator in $\mathcal{H}_i$. According to \eqref{aftermeas}, the state of the composite system after the measurement on $X_j$ correponding to $\tilde A^{\left(j\right)}_i$ is
\begin{equation}\label{aftermeascomp}
 \rho'_i=\left(\left(\tilde A^{\left(j\right)}_i\right)^{\ast}\rho\tilde A^{\left(j\right)}_i\right)/\left({\Tr\left(\left(\tilde A^{\left(j\right)}_i\right)^{\ast}\rho\tilde A^{\left(j\right)}_i\right)}\right).
\end{equation}
Consequently (cf. formula \eqref{partialtraces}), the state of the subsystem $X_k$ after the measurement corresponds to 
\begin{equation}\label{redaftermeascomp}
 \rho'^{\left(k\right)}_i=\Tr_1\ldots\Tr_{k-1}\Tr_{k+1}\ldots\Tr_n\left(\left(\left(\tilde A^{\left(j\right)}_i\right)^{\ast}\rho\tilde A^{\left(j\right)}_i\right)/\left({\Tr\left(\left(\tilde A^{\left(j\right)}_i\right)^{\ast}\rho\tilde A^{\left(j\right)}_i\right)}\right)\right).
\end{equation}
One can easily check that $\rho'^{\left(k\right)}_i$ does not need to be equal to $\rho^{\left(k\right)}$ when $\rho$ is not of the product form $\rho^{\left(1\right)}\otimes\ldots\otimes\rho^{\left(n\right)}$. To discuss this issue further, let us concentrate on bipartite systems, as we shall do in the remaining parts of the thesis. Let the composite system consist of two parts $A$ and $B$, whose states are described by elements of $M\left(V\right)$ and $M\left(U\right)$ (resp.), $V$ and $U$ being some finite-dimensional linear spaces. The state of the composite system is still denoted with $\rho$.
Let $\tilde A_i=A_i\otimes\mathbbm{1}$ ($i=1,\ldots,l$) and $\tilde B_j=\mathbbm{1}\otimes B_j$ ($j=1,\ldots,m$) be measurement operators pertaining to measurements on $A$ and $B$, respectively. Denote with $a_i$, $b_j$ the measurement results corresponding to $\tilde A_i$ and $\tilde B_j$.  Obviously, $\left[\tilde A_i,\tilde B_j\right]=0$ for all $i=1\ldots l$ and $j=1\ldots m$. Using the vanishing commutators, it is easy to obtain the following equalities,
\begin{equation}\label{fml.Pbj.1}
\fnc{P}{b_j}=\fnc{\Tr}{{\tilde B_j}^{\ast}\rho \tilde B_j}=\fnc{\Tr}{{\tilde B_j}^{\ast}\left(\sum_i{\tilde A_i}^{\ast}\rho\tilde A_i\right)\tilde B_j}
\end{equation}
and
\begin{equation}\label{fml.Pai.5}
\fnc{P}{a_i}=\fnc{\Tr}{{\tilde A_i}^{\ast}\rho \tilde A_i}=\fnc{\Tr}{{\tilde A_i}^{\ast}\left(\sum_j{\tilde B_j}^{\ast}\rho\tilde B_j\right)\tilde A_i}
\end{equation}
for the probabilities to obtain $b_j$ in a measurement on $B$ and the probability $P\left(a_i\right)$ to obtain $a_i$ in a measurement on $A$. Note that $\sum_i{\tilde A_i}^{\ast}\rho\tilde A_i$ ($\sum_j{\tilde B_j}^{\ast}\rho\tilde B_j$) is by definition \eqref{measnonselect} the state of the composite system after a nonselective measurement on its part $A$ ($B$, resp.). Thus the latter equalities in \fml{Pbj.1} and \fml{Pai.5} have an important physical content. As long as $A$ and $B$ are causally disconnected, the result of the measurement on $A$ cannot be known at the point of doing $B$ and vice versa. From ``the point of view of $B$'', a measurement on $A$ is nonselective as long as $B$ does not know the result $a_i$. In the nonselective scenario, we see from \fml{Pbj.1} that probabilities experienced in a measurement on part $B$ are equal $\fnc{\Tr}{{\tilde B_j}^{\ast}\rho\tilde B_j}$, no matter if a measurement on $A$ was done or not. Thus the reality at the point of doing $B$ changes only after the result obtained for $A$ becomes known to $B$, which must happen by some previously discovered communication channel (from \fml{Pai.5} we can get a similar conclusion with $A$ and $B$ interchanged). In other words, $A$ and $B$ cannot use quantum mechanics to increase their communication speed.  Hence we have explained that the ``instantaneous'' influence of $A$ on $B$ (and vice versa) in quantum mechanics is a matter of mathematical description. However, there is still something magical to it, even if it does not let things happen faster than the speed of light.

If we wanted to strip quantum mechanics of all its mystery, we should not only prove that the quantum mechanical decription of reality does not allow supraluminal signaling, but also that any correlations between the results obtained for $A$ and $B$ can be explained in a classical way. By this we mean that the correlations are set in advance, however there is more than a single possible way of setting them and we do not know which one is used in an individual experiment. For example, we have a white, a red and a black ball. With probability $\frac{1}{2}$, we give the red ball to $A$ and the white one to $B$. Otherwise, we give the red one to $B$ and the black one to $A$.   Similar things should amount for the randomness observed in quantum systems. In more mathematical terms, we would like to have a classical system with a corresponding probabilistic space $\left(S,\Sigma,\mu\right)$ such that every state $s\in\Sigma$ of the system corresponds to definite measurement outcomes $a_{i\left(s\right)}$ and $b_{j\left(s\right)}$. In such case, the probalilities to obtain $a_{i}$ and $b_{j}$ are given by the following formulas,
\begin{eqnarray}
\fnc{P}{a_i}=\int_S\fnc{\delta}{i,\fnc{i}{s}}d\fnc{\mu}{s}\label{fml.hiddenvar.1},\\
\fnc{P}{b_j}=\int_S\fnc{\delta}{j,\fnc{j}{s}}d\fnc{\mu}{s}\label{fml.hiddenvar.2},
\end{eqnarray}
where $\mu$ denotes the measure on $S$ and the $\delta$ function is $1$ if its two arguments are equal and it equals $0$ otherwise. We can also write a formula for the probability of observing $\left(a_{i},b_{j}\right)$ in a measurement on both $A$ and $B$,
\begin{equation}\label{fml.hiddenvar.3}
 \fnc{P}{a_i,b_j}=\int_S\fnc{\delta}{i,\fnc{i}{s}}\fnc{\delta}{j,\fnc{j}{s}}d\fnc{\mu}{s}.
\end{equation}
Equivalently, formulas \fml{hiddenvar.1}, \fml{hiddenvar.2} and \fml{hiddenvar.3} may be written as
\begin{eqnarray}
\fnc{P}{a_i}=\int_{S}\fnc{P_{\omega}}{a_i}d\fnc{\mu'}{\omega}\label{fml.hiddenvar.7},\\
\fnc{P}{b_j}=\int_{S}\fnc{P_{\omega}}{b_j}d\fnc{\mu'}{\omega}\label{fml.hiddenvar.8}\hskip 2 mm
\end{eqnarray}
and
\begin{equation}\label{fml.hiddenvar.9}
 \fnc{P}{a_i,b_j}=\int_{S}\fnc{P_{\omega}}{a_i}\fnc{P_{\omega}}{b_j}d\fnc{\mu'}{\omega},
\end{equation}
where we consider some other $\sigma$-algebra $\Sigma'\subset\Sigma$ on $S$ (with the induced measure $\mu'$) and adjust the functions $P_{\omega}$ accordingly.
In simple words, we can group the elementary events $s$ into bigger events $\omega$ if we prefer to retain some inherent randomness, expressed by $P_{\omega}$, in our classical states. Conceptual models of quantum mechanics where probabilities are calculated as in the formulas above are called {\it hidden variable models} because the states $s$ (or $\omega$) are assumed not to be known to physicist at the moment.

We may expect that $\fnc{P}{a_i}$, $\fnc{P}{b_j}$ and $\fnc{P}{a_i,b_j}$ are related somehow as a consequence of \fml{hiddenvar.1}, \fml{hiddenvar.2} and \fml{hiddenvar.3} (or \fml{hiddenvar.7}, \fml{hiddenvar.8} and \fml{hiddenvar.9}). Indeed, either set of formulas for $\fnc{P}{a_i}$, $\fnc{P}{b_j}$ and $\fnc{P}{a_i,b_j}$ can be used to prove that quantum mechanics is not equivalent to any hidden variable model. Inequalities on measurement probabilities in certain experiments can be obtained which do not always hold for the measurement probabilities predicted by quantum mechanics. Probably the most popular ones are the Bell and the CHSH inequalities  (see e.g. \cite{ref.Benetti}). Physical experiments strongly support the quantum mechanical way of calculating probabilities because the inequalities obtained for hidden variable models are violated. A discussion is still going on about closing possible loopholes in the experiments (see e.g. \cite{bellinek}), but there are no reasons to expect that hidden variable models may suffice for description of reality. This is often expressed by saying that quantum systems (like $A$ and $B$) can be {\it entangled}. If this is the case, the quantum state $\rho$ of the composite system is also called entangled.

Note that there is a huge family of quantum states which are not entangled. It is a simple exercise to check that all the states that satisfy
\begin{equation}\label{sepstates}
 \rho=\sum_ip_i\rho_i\otimes\sigma_i,
\end{equation}
for $p_i>0$, $\sum_ip_i=1$, $\rho_i\in M\left(V\right)$ and $\sigma_i\in M\left(U\right)$, admit a hidden variable description. We call these states {\it separable} \cite{ref.Werner}. All the other states are called {\it entangled} in the literature, but it should be kept in mind that this definition does not always imply entanglement in the sense discussed above.
It has been shown in \cite{ref.Werner} that there exist states admitting hidden variable description which are not of the form \eqref{sepstates}. However, the two definions seem to be very close to each other. Moreover, the importance of the distinction between separable and non-separable (entangled) states can be justified in a different way. Separable states are precisely the states which can be created from product states by so-called Local Operations and Classical Communication (LOCC) protocols.  Here, LOCC amounts to a local measurement on the first subsystem, followed by a local measurement on the second subsystem, with the second measurement conditioned on the result of the first measurement,
\begin{equation}
 \prj{\psi}\otimes\prj{\phi}\mapsto\sum_{i,j}\left(B^{\ast}_{ij}\otimes A^{\ast}_j\right)\left(\prj{\psi}\otimes\prj{\phi}\right)\left(B_{ij}\otimes A_j\right),
\end{equation}
where $\sum_iA_iA_i^{\ast}=\mathbbm{1}$ and $\sum_jB_{ij}B_{ij}^{\ast}=\mathbbm{1}\forall_j$. Entangled states cannot be created in the above way. This time the distinction is strict. 

\subsection{Quantum channels}\label{sec.Channels} Let us consider finite-dimensional $\mathcal{H}=V$, $\dim V=d$. We shall prove that nonselective measurements in the sense of the previous section are the only transformations of the type $\Phi:M\left(V\right)\ni\rho\mapsto\Phi\left(\rho\right)\in M\left(V\right)$ admissible in quantum mechanics. They are often called {\it quantum channels} to emphasize that they may correspond to elements like optical fibers, in fact used to send quantum states from one place to another. It should be kept in mind that there exist simple situations, e.g. when the system and its environment are entangled, where the knowledge of $\rho$ for the system is not sufficient to determine its future evolution. Then, one cannot describe the transformation of $\rho$ as a quantum channel \cite{Pachukas,Sudarshan,Buzek}. 
We concentrate on situations where such description is possible.

To prove our assertion, it is sufficient to observe that any quantum channel $\Phi$ must be a {\it completely positive} and {\it trace preserving} linear map\footnote{the fact that $\Phi$ must be linear can be proved as in \cite{pracamag}}. Let us denote with $\mathbbm{1}_k$ the identity operator on $L\left(\mathbbm{C}^k\right)$. Complete positivity of $\Phi$ means that the map $\Phi\otimes\mathbbm{1}_k$ maps positive operators $\rho$ into positive operators for arbitrary $k$. That is,
\begin{equation}\label{positivemap}
 \forall_{k\in\mathbbm{N}}\forall_{\rho\in P\left(V\otimes\mathbbm{C}^k\right)}\left(\Phi\otimes\mathbbm{1}_k\right)\rho\geqslant 0.
\end{equation}
In other words, complete positivity of $\Phi$ means that $\Phi\otimes\mathbbm{1}_k$ is a {\it positive map} for arbitrary $k\in\mathbbm{1}_k$. For further convenience, let us denote the set of completely positive maps of \LV with $\cpositivitymapsb{V}$. The condition that $\Phi$ must be completely positive can be justified in the following way. Consider a quantum system far away from the one described by $V$. Let the states of the other system correspond to elements of $M\left(\mathbbm{C}^k\right)$. It seems reasonable to assume that the systems can be chosen in such a way that they do not interact.   Nevertheless, it is conceivable that they have been prepared in an arbitrary state $\eta\in M\left(V\otimes\mathbbm{C}^k\right)$. Because of the lack of interaction, the fact that the first system is being sent through a quantum channel cannot affect the state of the second system. Consequently, it can be shown that the transformation of $\eta$ must be of the form $\eta\mapsto\left(\Phi\otimes\mathbbm{1}_k\right)\eta$. But $k$ and $\eta$ were arbitrary and $\left(\Phi\otimes\mathbbm{1}_k\right)\eta$ must turn up as a density matrix and thus positive. Hence condition \eqref{positivemap} must hold.

It follows from the Choi theorem on completely positive maps (Theorem \ref{Choithm} from Section \ref{sec.Jamiolkowski}) that any $\Phi$ that satisfies \eqref{positivemap} must be of the form
\begin{equation}\label{Krausrep}
 \Phi:\rho\mapsto\sum_{i=1}^lA^{\ast}_i\rho A_i
\end{equation}
for some operators $\left\{A_i\right\}_{i=1}^l\subset L\left(V\right)$. On the other hand, the trace preserving condition
\begin{equation}\label{Trprcond}
 \Tr\rho=\Tr\Phi\left(\rho\right)=\Tr\left(\sum_{i=1}^lA^{\ast}_i\rho A_i\right)=\Tr\left(\left(\sum_{i=1}^lA_iA^{\ast}_i\right)\rho\right)\,\forall_{\rho\in L\left(V\right)}
\end{equation}
implies that $\sum_{i=1}^lA_iA^{\ast}_i=\mathbbm{1}$. Condition \eqref{Krausrep} together with \eqref{Trprcond} imply that $\Phi$ is the nonselective measurement $\rho\mapsto\rho_{\times}$  corresponding to the measurement operators $\left\{A_i\right\}_{i=1}^l\subset L\left(V\right)$, with $\rho_{\times}$ given as in equation \eqref{measnonselect}. Thus we have proved our assertion.

The sum on the right hand side of \eqref{Krausrep} is a {\it Kraus representation} \cite{ref.Kraus} of $\Phi$. The operators $A_i$ are called {\it Kraus operators}. Note that there usually exists more than a single Kraus representation of a given $\Phi$.

\subsection{Jamio\l kowski isomorphism}\label{sec.Jamiolkowski}Let $V$ be a finite-dimensional linear space as in the previous section. Assume $U$ is finite-dimensional as well ($\dim U=h$).
The linear spaces $L\left(L\left(V\right),L\left(U\right)\right)$ and $L\left(U\otimes V\right)$ have the same dimension and thus are isomorphic. There exists an isomorphism between the two spaces called the {\it Jamio\l kowski isomorphism} which is especially suited for the purpose of testing complete positivity. It is defined in the following way,
\begin{equation}\label{Jamiolkowskiisomorphism}
J:\,L\left(L\left(V\right),\lu\right)\ni\Phi\mapsto\left(\Phi\otimes\mathbbm{1}\right)\left|\psi_+\right>\left<\psi_+\right|\in L\left(U\otimes V\right), 
\end{equation}
where $\mathbbm{1}$ denotes the identity operator on $L\left(V\right)$ and $\left|\psi_+\right>$ is the maximally entangled state on $V\otimes V$, $\left|\psi_+\right>=\frac{1}{\sqrt{d}}\sum_{\alpha=1}^d\left|\alpha\right>\left|\alpha\right>$, $\left\{\left|\alpha\right>\right\}_{\alpha=1}^d$ being an orthonormal basis of $V$. Let us also introduce an orthonormal basis $\left\{\left|a\right>\right\}_{a=1}^h$ of $U$.
In index notation, the action of $J$ amounts to swapping a pair of indices,
\begin{equation}\label{reshuffling}
 \Phi_{ab,\gamma\delta}\rightarrow\left(J\left(\Phi\right)\right)_{a\gamma,b\delta}=\Phi_{ab,\gamma\delta}
\end{equation}
where $\Phi_{ab,\gamma\delta}$ are matrix elements of $\Phi$ w.r.t. the bases $\left\{\pprj{a}{b}\right\}_{a,b=1}^{h}$, $\left\{\pprj{\gamma}{\delta}\right\}_{\gamma,\delta=1}^{d}$ and $\left(J\left(\Phi\right)\right)_{a\gamma,b\delta}$ are matrix elements of $J\left(\Phi\right)$ w.r.t. the basis $\left\{\left|\beta\right>\left|\delta\right>\right\}_{b,\delta=1}^{b=h,\delta=d}$.  We call the matrix operation \eqref{reshuffling} {\it reshuffling}.

Note that $\Phi$ in \eqref{Jamiolkowskiisomorphism} represents an arbitrary linear map from $L\left(V\right)$ to $\lu$, which does not have to be completely positive. For completely positive $\Phi$, we have the following theorem \cite{ref.Choi}.
\begin{theorem}[Choi]\label{Choithm}
 Let $\Phi$ be a linear map in $L\left(V\right)$. Let $J$ be defined as in \eqref{Jamiolkowskiisomorphism} (with $U=V$). Then we have
\begin{equation}\label{Choithmequiv}
 \Phi\text{ is completely positive}\Longleftrightarrow\,J\left(\Phi\right)\geqslant 0.
\end{equation}
\qed
\end{theorem}
\noindent That is, completely positive maps $\Phi$ correspond to positive operators $J\left(\Phi\right)$. This allows to check complete positivity of arbitrary $\Phi$. The fact that any completely positive map $\Phi$ can be written in the Kraus form \eqref{Krausrep} follows from the spectral decomposition of $\left(\Phi\otimes\mathbbm{1}\right)\left|\psi_+\right>\left<\psi_+\right|$,
\begin{equation}\label{specdec}
 \left(\Phi\otimes\mathbbm{1}\right)\left|\psi_+\right>\left<\psi_+\right|=\sum_i\lambda_i\left|\alpha_i\right>\left<\alpha_i\right|,
\end{equation}
where $\lambda_i>0$ for all $i$ and $\alpha_i$ are elements of an orthonormal basis of $U\otimes V$. We leave as an excercise for the reader to prove that the representation \eqref{specdec} of $\left(\Phi\otimes\mathbbm{1}\right)\left|\psi_+\right>\left<\psi_+\right|$ implies that $\Psi$ can be written in the Kraus form \eqref{Krausrep} with operators $A_i$ such that the matrix elements $\left(A_i\right)_{b\delta}$ of $A_i$ w.r.t. $\left\{\left|b\right>\left|\delta\right>\right\}_{b,\delta=1}^{b=h,\delta=d}$ satisfy
\begin{equation}\label{alphacoords}
 \left|\alpha_i\right>=\sum_{b,\delta=1}^{h,d}\left(A_i\right)_{b\delta}\left|b\right>\left|\delta\right>.
\end{equation}

In the case $U=V$, note that $J$ is not a homomorphism of the multiplicative structures of $L\left(L\left(V\right)\right)$ and $L\left(V\otimes V\right)$. In other words, $J\left(\Phi\Psi\right)=J\left(\Phi\right)J\left(\Psi\right)$ does not hold in general for $\Psi,\Phi\in L\left(L\left(V\right)\right)$. However, it is possible to define an alternative multiplicative structure $\left(L\left(V\otimes V\right),\odot\right)$ in such a way that $J$ is a homomorphism between $L\left(L\left(V\right)\right)$ and $\left(L\left(V\otimes V\right),\odot\right)$.
For $A,B\in L\left(V\otimes V\right)$, we define \cite{ref.Roga}
\begin{equation}\label{circmul}
 A\odot B=J\left(J^{-1}\left(A\right)J^{-1}\left(B\right)\right).
\end{equation}
It is now obvious that $J\left(\Phi\Psi\right)=J\left(\Phi\right)\odot J\left(\Psi\right)$ for arbitrary $\Psi,\Phi\in L\left(L\left(V\right)\right)$.

To see how $\odot$ looks like in index notation, let $A$ and $B$ be elements of $L\left(V\otimes V\right)$ with matrix elements $A_{\alpha\beta,\gamma\delta}$ and $A_{\alpha\beta,\gamma\delta}$, resp.  Let $\left(A\odot B\right)_{\alpha\beta,\gamma\delta}$ be the matrix elements of $A\odot B$. From the definitions of $\odot$ and $J$ (formulas \eqref{Jamiolkowskiisomorphism} and \eqref{circmul} in Section \ref{sec.Jamiolkowski}) we have
\begin{equation}\label{fml.altprod.1}
 \left(A\odot B\right)_{\alpha\beta,\gamma\delta}=\fnc{J}{\fnc{J^{-1}}{A}\fnc{J^{-1}}{B}}_{\alpha\beta,\gamma\delta}=\left(\fnc{J^{-1}}{A}\fnc{J^{-1}}{B}\right)_{\alpha\gamma\beta\delta}.
\end{equation}
The same as $J$, $J^{-1}$ corresponds to reshuffling indices. Thus we get
\begin{equation}\label{fml.altprod.2}
\left(\fnc{J^{-1}}{A}\fnc{J^{-1}}{B}\right)_{\alpha\beta,\gamma\delta}=\sum_{\xi,\zeta}\fnc{J^{-1}}{A}_{\alpha\gamma\xi\zeta}\fnc{J^{-1}}{B}_{\xi\zeta\beta\delta}=\sum_{\xi,\zeta}A_{\alpha\xi,\gamma\zeta}B_{\xi\beta,\zeta\delta}.
\end{equation}
From \fml{altprod.1} and \fml{altprod.2} we have the formula
\begin{equation}\label{fml.altprod.3}
\left(A\odot B\right)_{\alpha\beta,\gamma\delta}=\sum_{\xi,\zeta}A_{\alpha\xi,\gamma\zeta}B_{\xi\beta,\zeta\delta}.
\end{equation}
As we see from \fml{altprod.3}, $\odot$ is different from the standard product of operators. It is also easy to notice that the operation given by \fml{altprod.3} depends on the choice of basis of $V$. This should be expected since $J$ is basis-dependent. We shall call $\odot$ {\it circled product} in $L\left(V\otimes V\right)$.

\section{Convex cone dualities}\label{sec.ConeDual}
We explained in Section \ref{sec.Channels} that only completely positive maps  may be considered as corresponding to physical processes. However, a wider class of {\it positive maps} proves to be useful in testing separability of quantum states (cf. Corollary \ref{prop.kposmapscr} below for $k=1$). Its characterization has also been a long-standing problem in pure mathematics \cite{ref.Stormer}. In this section, we discuss classes of maps of \LV  that are positive, but not necessarily completely positive, as well as classes of maps that fulfil even stronger properties than complete positivity. The wider classes turn out to be related to the narrower ones by a convex duality relation that we describe below. Moreover, in Section \ref{sec.genposmapscr} we show that they all need to satisfy significantly stronger conditions, which are generalizations of the positive maps criterion by Horodeccy \cite{ref.Horodeccy}.
Recent St\o rmer's work \cite{St09symmetric} gives a further generalization, which may prove to be the maximum possible one.


\subsection{Geometry in operator spaces}\label{sec.GeoOp}
As in the previous sections, let $V$ be a linear space of dimension $d<+\infty$ with an inner product $\hprod{.}{.}$. The operator space $\lv$ is naturally endowed with a {\it Hilbert-Schmidt inner product},
\begin{equation}\label{fml.HSProddef}
 \left(A|B\right)=\Tr\left(A^{\ast}B\right),
\end{equation}
where $\ast$ denotes the adjoint of an operator w.r.t. $\hprod{.}{.}$. Because $\vov$ is also a finite-dimensional vector space equipped with an inner product, the same construction works for $\lvov$. Moreover, we can iteratively use it for $\fnc{L}{\lv}$ or even for $\fnc{L}{\fnc{L}{\lv}}$ a.s.o. Thus we have well-defined inner products and adjoint operations in all these spaces. It turns out that the Jamio\l kowski isomorphism defined in Section \ref{sec.Jamiolkowski} is an isometry of $\fnc{L}{\lv}$ and $\lvov$.
\begin{proposition}\label{prop.Jisometry}Let the spaces $\lvov$ and $\fnc{L}{\lv}$ be equipped with the inner products following from the construction above. The  isomorphism $J$ defined by formula \eqref{Jamiolkowskiisomorphism} is an isometry between $\lvov$ and $\fnc{L}{\lv}$, i.e.
\begin{equation}\label{isome}
 \left(\Phi|\Psi\right)=\left(\fnc{J}{\Phi}|\fnc{J}{\Psi}\right)
\end{equation}
for arbitrary $\Phi,\Psi\in\fnc{L}{\lv}$.
\begin{proof}
 Let us first observe that $\prj{\psi_+}=\sum_{\alpha,\beta=1}^d\pprj{\alpha}{\beta}\otimes\pprj{\alpha}{\beta}$. Thus by definitions \eqref{Jamiolkowskiisomorphism} and \eqref{fml.HSProddef}, we have
\begin{equation}\label{prodHSright}
 \left(\fnc{J}{\Phi}|\fnc{J}{\Psi}\right)=\sum_{\alpha,\beta=1}^d\sum_{\alpha',\beta'=1}^d\Tr\left(\left(\Phi\left(\pprj{\alpha}{\beta}\right)\otimes\pprj{\alpha}{\beta}\right)^{\ast}\Psi\left(\pprj{\alpha'}{\beta'}\right)\otimes\pprj{\alpha'}{\beta'}\right).
\end{equation}
This is the same as
\begin{multline}\label{prodHSright2}
 \sum_{\alpha,\beta=1}^d\sum_{\alpha',\beta'=1}^d\Tr\left(\left(\Phi\left(\pprj{\alpha}{\beta}\right)\otimes\pprj{\alpha}{\beta}\right)^{\ast}\Psi\left(\pprj{\alpha'}{\beta'}\right)\otimes\pprj{\alpha'}{\beta'}\right)=\\
=\sum_{\alpha,\beta=1}^d\Tr\left(\Phi\left(\pprj{\alpha}{\beta}\right)^{\ast}\Psi\left(\pprj{\alpha}{\beta}\right)\right),
\end{multline}
where we used the simple fact $\Tr\left(\left(X\otimes Y\right)\left(X'\otimes Y'\right)\right)=\Tr\left(XX'\right)\Tr\left(YY'\right)$, as well as $\Tr\left(\left(\pprj{\alpha}{\beta}\right)^{\ast}\pprj{\alpha'}{\beta'}\right)=\delta_{\alpha\alpha'}\delta_{\beta\beta'}$. On the other hand, we have
\begin{multline}\label{prodHSleft}
 \left(\Phi|\Psi\right)=\Tr\left(\Phi^{\ast}\Psi\right)=\sum_{\alpha,\beta=1}^d\bigl(\pprj{\alpha}{\beta}|\Phi^{\ast}\Psi\left(\pprj{\alpha}{\beta}\right)\bigr)=\\
=\sum_{\alpha,\beta=1}^d\bigl(\Phi\left(\pprj{\alpha}{\beta}\right)|\Psi\left(\pprj{\alpha}{\beta}\right)\bigr)=\sum_{\alpha,\beta=1}^d\Tr\left(\Phi\left(\pprj{\alpha}{\beta}\right)^{\ast}\Psi\left(\pprj{\alpha}{\beta}\right)\right),
\end{multline}
which is the same as the last expression in \eqref{prodHSright2}.
\end{proof}
\end{proposition}
In contrast to the proposition above, $J$ is not a $\ast$-morphism between $\lvov$ and $\fnc{L}{\lv}$, i.e. $J\left(\Phi^{\ast}\right)\neq J\left(\Phi\right)^{\ast}$ in general. Therefore, self-adjoint operators belonging to $\fnc{L}{\lv}$ are not mapped into self-adjoint elements of $\lvov$. In the following, we shall be interested only in self-adjoint operators in $\vov$ and the corresponding maps of $\lv$. In other words, we restrict ourselves to $\hvov$ and $J^{-1}\left(\hvov\right)$. This is motivated by the fact that all positive maps of $\lv$ (cf. Section \ref{sec.kpospsposdual}) are mapped by $J$ into self-adjoint operators in $\vov$.
Since $J$ is not a $\ast$-morphism, $J^{-1}\left(\hvov\right)\neq\fnc{H}{\lv}$. However, just like $\hvov$, $J^{-1}\left(\hvov\right)$ is a linear space over $\setR$. In addition to that, the Hilbert-Schmidt product in $\fnc{L}{\lv}$ is symmetric when we restrict it to $J^{-1}\left(\hvov\right)$. This follows simply from Proposition \ref{prop.Jisometry} and the fact that $\left(A|B\right)=\left(B|A\right)=\Tr\left(AB\right)$ for $A,B\in\hvov$. Consequently, we may define the following {\it duality relation}
\begin{equation}\label{dualityrefhvov}
\mathcal{A}^{\circ}=\left\{B\in\hvov|\left(A|B\right)\geqslant 0\,\forall_{A\in\hvov}\right\}
\end{equation}
between subsets $\mathcal{A}$, $\mathcal{A}^{\circ}$ of $\hvov$ or an analogous relation
\begin{equation}\label{dualityrefJ-1hvov}
\mathcal{F}^{\circ}=\left\{\Psi\in J^{-1}\left(\hvov\right)|\left(\Phi|\Psi\right)\geqslant 0\,\forall_{\Phi\in J^{-1}\left(\hvov\right)}\right\}
\end{equation}
between $\mathcal{F},\mathcal{F}^{\circ}\subset J^{-1}\left(\hvov\right)$. We call $\mathcal{X}^{\circ}$ the {\it dual} of $\mathcal{X}$, no matter if $\mathcal{X}$ is a subset of $J^{-1}\left(\hvov\right)$ or of $\hvov$. It should be noted that $\left(\Phi|\Psi\right)\neq\Tr\left(\Phi\Psi\right)$ for $\Phi,\Psi\in J^{-1}\left(\hvov\right)$, but the definition \eqref{dualityrefJ-1hvov} is correct because of the mentioned symmetry of $\left(.|.\right)$ in $J^{-1}\left(\hvov\right)$. From Proposition \ref{prop.Jisometry} we immediately obtain
\begin{equation}\label{Jiso_duality}
 J\left(\mathcal{F}^{\circ}\right)=J\left(\mathcal{F}\right)^{\circ}
\end{equation}
for any subset $\mathcal{F}$ of $J^{-1}\left(\hvov\right)$. 

Note that the dual is always a convex cone, i.e. $p X+q Y\in\mathcal{F}^{\circ}$ for arbitrary $X,Y\in\mathcal{F}^{\circ}$ and $p,q\in\left[0;+\infty\right)$. In the case where $\mathcal{F}$ is also a convex cone, it is natural to ask about the relation between $\mathcal{F}$ and $\mathcal{F}^{\circ\circ}$. One has (cf. \cite{pracamag,ref.Rockafellar})
\begin{proposition}\label{prop.dbldual}For any convex cone $\mathcal{F}$ in $\hvov$ (or in $J^{-1}\left(\hvov\right)$, one has
\begin{equation}\label{dbldualeq1}
 \mathcal{F}^{\circ\circ}=\overline{\mathcal{F}}.
\end{equation}
In particular, for closed convex cones $\mathcal{F}^{\circ\circ}=\mathcal{F}$.\qed
\end{proposition}


\subsection{Duality between k-positive and k-superpositive maps}\label{sec.kpospsposdual}
Positivity and $k$-positivity conditions for maps have already been mentioned in Section \ref{sec.Channels}, but let us make them more explicit. We call a map $\Phi\in\fnc{L}{\lv}$ {\it positive} iff it fulfils
\begin{equation}\label{fml.PosCond}
 \forall_{\rho\in \pv}\fnc{\Phi}{\rho}\geqslant 0,
\end{equation}
 i.e. when $\Phi$ maps positive operators in \LV into positive operators. If the map $\Phi\otimes\mathbbm{1}_k$ of $V\otimes\setC^k$ (for $k\in\setN$) is positive, we say that $\Phi$ is {\it $k$-positive}. For further convenience, let us denote the set of positive maps of \LV with $\positivitymapsb{V}$ and the set of $k$-positive maps with $\kpositivitymapsb{k}{V}$.
It is clear from \eqref{positivemap} that a map of \LV is completely positive iff it is $k$-positive for arbitrary $k$, which is the usual way to describe complete positivity. It has been known for decades \cite{Jamiolkowski1} that positive maps are related by the Jamio\l kowski isomorphism to so-called {\it block positive} operators in $\vov$. These are the elements $A$ of \LVoV that satisfy
\begin{equation}\label{bposdef}
 \hprod{u\otimes v}{\fnc{A}{u\otimes v}}\geqslant 0
\end{equation}
for all $u,v\in V$. That is, block positive operators are positive on product vectors in $\vov$.  We can now make explicit the relation between positive maps and block positive operators \cite{Jamiolkowski1},
\begin{proposition}[Jamio\l kowski]\label{Jamiolkowskithm}
\begin{equation}\label{Jamiolkowskithmeq}
 J\left(\positivitymapsb{V}\right)=\bposopsb{\vov},
\end{equation}
where $\bposopsb{\vov}$ denotes the set of operators in \VoV that satisfy \eqref{bposdef}.\qed
\end{proposition}
The condition \eqref{bposdef} implies Hermiticity, which can be proved as in \cite{pracamag}. Therefore $\bposopsb{\vov}$ is a subset of $\hvov$ and thus $\positivitymapsb{V}$ is a subset of $J^{-1}\left(\hvov\right)$, as we mentioned in the previous section.

It is not difficult to prove \cite{SSZ09} a more general statement than Proposition \ref{Jamiolkowskithm}, concerning $k$-positive maps and their relation to the so-called set of \textit{$k$-block positive operators} 
($k\in\setN$),
\begin{equation}\label{kblockposdef}
 \kbposopsb{k}{V\otimes V}:=
\left\{A\,\vline\innerpr{\sum_{i=1}^k u_i\otimes v_i}
{A\sum_{l=1}^k u_l\otimes v_l}\geqslant 0\,
\forall_{\seq{u_i}{i=1}{k},
\seq{v_i}{l=1}{k}\subset V}\right\},
\end{equation}
where the $A$'s are elements of $\fnc{L}{ V \otimes V }$. 
Note that $\kbposopsb{1}{ V \otimes V }=\bposopsb{ V \otimes V }$ and $\kbposopsb{d}{ V \otimes V }=\posopsb{ V \otimes V }$.
We have for arbitrary $k\in\setN$ the following 
\begin{proposition}\label{prop.kPoskBP}
\begin{equation}\label{kbpkposJ}
 J\left(\kpositivitymapsb{k}{V}\right)=\kbposopsb{k}{\vov}.
\end{equation}
\qed
\end{proposition}
In particular, taking $k\geqslant d$, one recovers the Choi's theorem (Theorem \ref{Choithm}).
The above result appeared already in \cite{TT83} and was also the subject of the paper \cite{ref.Ranade}. It is clear from definition \eqref{kblockposdef} that the sets $\kbposopsb{k}{V}$ form a chain of subsets, $\bposopsb{\vov}\supset\kbposopsb{2}{\vov}\supset\ldots\supset\kbposopsb{d}{\vov}=\ldots=\posopsb{\vov}$, where the equalities at the end follow from $\kbposopsb{d}{\vov}=\posopsb{\vov}$. In the same way, $\positivitymapsb{V}\supset\ldots\supset\kpositivitymapsb{d}{V}=\ldots=\cpositivitymapsb{V}$. 

We will also be interested in a family of subsets of $\cpositivitymapsb{V}$, called {\it$k$-superpositive} maps ($k=1,\ldots,d$). For a given $k$, they are defined in the following way,
\begin{equation}\label{kspdef}
 \kspositivitymapsb{k}{S}:=\left\{\Phi\,\vline\,\Phi:\xi\mapsto\sum_iX_i^{\ast}\xi X_i,\,X_i\in\lv,\rk X_i\leqslant k\,\forall_i\right\},
\end{equation}
where $\Phi$ refers to an element of $\fnc{L}{\lv}$. The maps in $\kspositivitymapsb{1}{V}$ are called {\it superpositive} \cite{An04} and will be denoted with $\spositivitymapsb{V}$ for simplicity. In the case when an additional trace-preserving condition is imposed,
superpositive maps are often called \textit{entanglement breaking channels} after the work by Horodecki, Shor and Ruskai \cite{HSR03}. Obviously, $\spositivitymapsb{V}\subset\kspositivitymapsb{2}{V}\subset\ldots\subset\kspositivitymapsb{d}{V}\subset\ldots$ . Actually, $\kspositivitymapsb{k}{V}=\cpositivitymapsb{V}$ for $k\geqslant d$ as a consequence of Choi's theorem and the representation of positive maps that we mentioned in Section \ref{sec.Channels}.

It is not difficult to find (cf. \cite{SSZ09} or \cite{pracamag}) the image of $\kspositivitymapsb{k}{V}$ by $J$,
\begin{proposition}
\label{propkSPkSep}
 Let $k$ be a positive integer.
 Let us define the set of \textit{$k$-entangled operators} on 
$ V \otimes V $,  
\begin{equation}\label{ksepdef}
 \ksepopsb{k}{ V \otimes V }:=
\convhull{\left\{\sum_{i,j=1}^k\diad{u_i\otimes v_i}{u_j\otimes v_j}\,
\vline\seq{u_i}{i=1}{k},\seq{v_j}{j=1}{k}\subset V \right\}}.
\end{equation}
The set of $k$-superpositive maps is isomorphic to 
$\ksepopsb{k}{ V \otimes V }$,
\begin{equation}
\label{ksepisokSP}
 \fnc{J}{\kspositivitymapsb{k}{ V }}=\ksepopsb{k}{ V \otimes V }.
\end{equation}
\qed
\end{proposition}
Note that $\ksepopsb{k}{\vov}$ is the same as the set of operators with Schmidt number less than or equal to $k$ \cite{ref.Ranade,ref.Zyczkowski,ref.Terhal}. In particular, $\ksepopsb{1}{\vov}$ equals the set of separable states on $\vov$. Is is now easy to notice the following
\begin{proposition}\label{kBPdualkSep}
\begin{equation}\label{eq1dualitykBPkSep}
 \kbposopsb{k}{V}=\left(\ksepopsb{k}{V}\right)^{\circ}.
\end{equation}
\begin{proof}
Follows directly from the definitions \eqref{kblockposdef}, \eqref{ksepdef} and the simple relation
\begin{equation}\label{relTrInner}
 \innerpr{\sum_{i=1}^k u_i\otimes v_i}
{A\sum_{j=1}^k u_j\otimes v_j}=
\Trb{A\sum_{i,j=1}\diad{u_j\otimes v_j}{u_i\otimes v_i}},
\end{equation}
where $A\in\lv$.
\end{proof}
\end{proposition}
It is not difficult to show (cf. \cite{pracamag}) that $\ksepopsb{k}{V}$ is a closed convex cone for arbitrary $k\in\setN$. Thus we can apply Proposition \ref{prop.dbldual} to \eqref{eq1dualitykBPkSep} and obtain
\begin{proposition}\label{prop.kSepdualkBP}
\begin{equation}
 \ksepopsb{k}{V}=\left(\kbposopsb{k}{V}\right)^{\circ}.
\end{equation}
\begin{proof}
 Follows directly from Propositions \ref{prop.dbldual} and \ref{kBPdualkSep}.
\end{proof}
\end{proposition}
Since the relation \eqref{Jiso_duality} holds, we can also formulate analogues of Propositions \ref{prop.kSepdualkBP} and \ref{kBPdualkSep} for maps of $\lv$,
\begin{proposition}\label{kposdualkspos}
\begin{equation}
 \kpositivitymapsb{k}{V}=\left(\kspositivitymapsb{k}{V}\right)^{\circ}.
\end{equation}
\begin{proof}
 A simple consequence of equality \eqref{Jiso_duality} and Propositions \ref{prop.kPoskBP}, \ref{propkSPkSep} \& \ref{kBPdualkSep}.
\end{proof}
\end{proposition}
\begin{proposition}\label{ksposdualkpos}
 \begin{equation}
 \kspositivitymapsb{k}{V}=\left(\kpositivitymapsb{k}{V}\right)^{\circ}.
\end{equation}
\begin{proof}
 Follows from \eqref{Jiso_duality} and Propositions \ref{prop.kPoskBP}, \ref{kBPdualkSep} \& \ref{prop.kSepdualkBP}.
\end{proof}
\end{proposition}
Let us conclude with the observation that all the convex cones in $\fnc{L}{\lv}$ discussed in the present section are closed under taking adjoints. Being more verbose, we can state the following
\begin{proposition}\label{prop.takingadjoints}Let $\mathcal{F}$ be any of the sets $\kpositivitymapsb{k}{V}$, $\kspositivitymapsb{k}{V}$ ($k=1,\ldots,d$). Define
\begin{equation}\label{def.adjointF}
 \mathcal{F}^{\ast}:=\left\{\Phi^{\ast}\,\vline\,\Phi\in\mathcal{F}\right\},
\end{equation}
where the adjoint is defined w.r.t. the Hilbert-Schmidt product in $\fnc{L}{\lv}$ (cf. beginning of Section \ref{sec.GeoOp}). We have
\begin{equation}\label{FeqFast}
 \mathcal{F}^{\ast}=\mathcal{F}.
\end{equation}
\begin{proof}Let $\Phi$ be an element of $\kspositivitymapsb{k}{ V }$ and $\Psi$ 
an element of $\kpositivitymapsb{k}{ V }$. We want to prove that $\hconj{\Phi}\in\kspositivitymapsb{k}{ V }$ and 
$\hconj{\Psi}\in\kpositivitymapsb{k}{ V }$.
Just as $\posopsb{ V \otimes V }$, the set 
$\posopsb{\setC^k\otimes V }$ is self-dual. Thus we have that
$\xi\in\posopsb{\setC^k\otimes V }\Leftrightarrow\Trb{\hconj{\xi}\zeta}
\geqslant 0\,\forall_{\zeta\in\posopsb{\setC^k\otimes V }}$. 
The definition of $k$-positivity of $\Psi$ can be restated as
\begin{equation}
\label{defkpos2}
 \Trb{\hconjb{\left(\identitymapn{k}\otimes\Psi\right) \xi}\zeta}\geqslant 0\,
\forall_{\xi,\zeta\in\posopsb{\setC^k\otimes V }}.
\end{equation}
Equivalently,
\begin{equation}\label{defkpos3}
 \Trb{\hconjb{\left(\identitymapn{k}\otimes\hconj{\Psi}\right) \zeta}\xi}\geqslant0\,
\forall_{\xi,\zeta\in\posopsb{\setC^k\otimes V }}.
\end{equation}
But this is just the condition \eqref{defkpos2} for $\hconj{\Psi}$. 
Hence $\Psi\in\kpositivitymapsb{k}{ V }\Leftrightarrow\hconj{\Psi}\in
\kpositivitymapsb{k}{ V }$. To prove an analogous equivalence for $\Phi$, 
it is enough to consider the specific case $\Phi:\xi\mapsto \hconj{X}\xi X$ with $\rk X\leqslant k$. 
We have
\begin{equation}
\label{adjointPhi}
 \left(\Phi\left(\xi\right)|\zeta\right)=\Trb{\hconjb{\hconj{X}\xi X}\zeta}=
\Trb{\hconj{\xi}\left(X\zeta\hconj{X}\right)}=\left(\xi|X\zeta\hconj{X}\right)
\end{equation}
This gives us $\hconj{\Phi}:\xi\mapsto X\xi\hconj{X}$. 
The ranks of $X$ and $\hconj{X}$ are equal, so
$\Phi\in\kspositivitymapsb{k}{ V }\Leftrightarrow\hconj{\Phi}\in
\kspositivitymapsb{k}{ V }$. This finishes the proof of the proposition.
\end{proof}
\end{proposition}
Note that Proposition \ref{prop.takingadjoints} does not mean that all the cones $\kspositivitymapsb{k}{V}$, $\kpositivitymapsb{k}{V}$ consist of self-adjoint maps $\Phi$ (remember $J^{-1}\left(\hvov\right)\neq\fnc{H}{\lv}$). The equality between $\mathcal{F}$ and $\mathcal{F}^{\ast}$ does not imply $\Phi^{\ast}=\Phi$ for all $\Phi$ in $\mathcal{F}$.


\subsection{Generalized positive maps criterion}\label{sec.genposmapscr}
The following proposition will be crucial for proving Theorem \ref{characterizationsksp}, which is the main result of Section \ref{sec.ConeDual}.

\begin{proposition}
\label{prodthmkSPkP} For all $\Phi\in\kspositivitymapsb{k}{S}$ and $\Psi\in\kpositivitymapsb{k}{V}$, we have
\begin{equation}\label{eqSPkPkclosed}
 \Phi\Psi\subset\kspositivitymapsb{k}{V}\quad\textnormal{and}\quad\Psi\Phi\subset\kspositivitymapsb{k}{V}.
\end{equation}
\begin{proof}
 We want to prove that
 $\Phi\Psi\in\kspositivitymapsb{k}{\HilbertSp}$ and 
$\Psi\Phi\in\kspositivitymapsb{k}{\HilbertSp}$ for 
arbitrary $k\in\setN$, whenever $\Phi\in\kspositivitymapsb{k}{\HilbertSp}$ and 
$\Psi\in\kpositivitymapsb{k}{\HilbertSp}$. It is sufficient to show this for 
$\Phi:\xi\mapsto\hconj{X}\xi X$ with an arbitrary $X\in\linearopsb{\HilbertSp}$ of rank $\leqslant k$. 
We prove first that $\Psi\Phi$ is an element of 
$\kspositivitymapsb{k}{\HilbertSp}$. For this we shall need the following lemma
\begin{lemma}
\label{lemmaPhikposofdiad}
 Let $\Psi\in\linearmapsb{\HilbertSp}$ be $k$-positive. 
For any $k$-element set of vectors $\seq{u_i}{i=1}{k}$, 
there exists $m\in\setN$ and vectors $\seq{w^{\left(n\right)}_l}{l,n=1}{l=k,n=m}\subset\HilbertSp$ 
such that
\begin{equation}
\label{Phiofadiad}
 \Psi\left(\diad{ u_i}{ u_j}\right)=\sum_{n=1}^m\diad{ w^{\left(n\right)}_i}{ w^{\left(n\right)}_j}
\end{equation}
for all $i,j\in\left\{1,\ldots,k\right\}$.
\begin{proof}
The operator $\left[\Psi\left(\diad{ u_i}{ u_j}\right)\right]_{i,j=1}^k$ belongs to
 $\linearopsb{\setC^k\otimes\HilbertSp}$. Since $ u$ is positive, 
$\left[\Psi\left(\diad{ u_i}{ u_j}\right)\right]\in\posopsb{\setC^k\otimes\HilbertSp}$, 
hence is a sum of positive rank $1$ operators, which are necessarily of the form 
$\left[\diad{ w_i^{\left(n\right)}}{ w_j^{\left(n\right)}}\right]_{i,j=1}^k$ with
 $\left\{ w_l^{\left(n\right)}\right\}_{l,n=1}^{l=k,n=m}$ as in the statement of the theorem.
\end{proof}
\end{lemma}
Now we can prove that $\Psi\Phi\in\kspositivitymapsb{k}{\HilbertSp}$. 
Let us take an arbitrary element $\xi\in\linearopsb{\HilbertSp}$. 
The fact that $\rk X\leqslant k$ is equivalent to 
$X=\sum_{i=1}^k\diad{ v_i}{ u_i}$ for some vectors $\seq{ v_i}{i=1}{k},
\seq{ u_j}{j=1}{k}\subset\HilbertSp$. Thus we get 
\begin{equation}
\label{formofAdmap}
 \Phi\left(\xi\right)=\sum_{i,j=1}^k\innerpr{ v_i}{\xi v_j}
\diad{ u_i}{ u_j}.
\end{equation}
Now we calculate the action of $\Psi\Phi$ on $\xi$,
\begin{equation}\label{actionofPhiAd}
 \Psi\Phi\left(\xi\right)=\sum_{i,j=1}^k\innerpr{ v_i}{\xi v_j}
\Psi\left(\diad{ u_i}{ u_j}\right)=\sum_{l=1}^m\sum_{i,j=1}^k\innerpr{ v_i}
{\xi v_j}\diad{ w^{\left(l\right)}_i}{ w^{\left(l\right)}_j}.
\end{equation}
This is a sum of terms of the form \eqref{formofAdmap} and 
we get $\Psi\Phi=\sum_{l=1}^m\Phi_l$, where the 
operators $\Phi_l:\xi\mapsto\sum_{i,j=1}^k|w^{\left(l\right)}_j \rangle\langle v_j|\xi| v_j\rangle\langle w^{\left(l\right)}_j|$ all 
belong to $\kspositivitymapsb{k}{V}$. Thus we have proved 
$\Psi\Phi\in\kspositivitymapsb{k}{\HilbertSp}$ in the case $\Phi:\xi\mapsto\hconj{X}\xi X$, 
which implies that $\Psi\Phi\in\kspositivitymapsb{k}{\HilbertSp}$ 
for arbitrary $\Phi\in\kspositivitymapsb{k}{\HilbertSp}$. 
We still need to show that $\Phi\Psi\in\kspositivitymapsb{k}{\HilbertSp}$. 
By Proposition \ref{prop.takingadjoints}, $\Phi\Psi\in\kspositivitymapsb{k}{\HilbertSp}$ 
is equivalent to $\hconjb{\Phi\Psi}=\hconj{\Psi}\hconj{\Phi}
\in\kspositivitymapsb{k}{\HilbertSp}$. The last equality holds according to
 Proposition \ref{prop.takingadjoints} and to the first part of the proof.
\end{proof}
\end{proposition}
\noindent In short, we proved that for any $\Phi$ $k$-superpositive and 
$\Psi$ $k$-positive, the products $\Phi\Psi$ and $\Psi\Phi$ are 
$k$-superpositive. Now we can prove the main result of Section \ref{sec.ConeDual}.
\begin{theorem}\label{characterizationsksp}
Let $\Phi\in J^{-1}\left(\hvov\right)$ and $k\in\setN$. The following 
conditions are equivalent:
\begin{enumerate}[1)]
\item $\Phi\in\kspositivitymapsb{k}{\HilbertSp}$,
\item $\Psi\Phi\in\kspositivitymapsb{k}{\HilbertSp}\,
\forall_{\Psi\in\kpositivitymapsb{k}{\HilbertSp}}$,
\item $\Psi\Phi\in\cpositivitymapsb{\HilbertSp}\,
\forall_{\Psi\in\kpositivitymapsb{k}{\HilbertSp}}$,
\item $\Trb{\proj{\psi_+}\left(\identitymap\otimes
\Psi\Phi\right)\left(\proj{\psi_+}\right)}\geqslant 0\,\,
\forall_{\Psi\in\kpositivitymapsb{k}{\HilbertSp}}$.
\end{enumerate}
\begin{proof}
$1)\Rightarrow 2)$ As we know from Proposition \ref{prodthmkSPkP}, 
$\Psi\Phi\in\kspositivitymapsb{k}{\HilbertSp}$ for 
$\Psi\in\kpositivitymapsb{k}{\HilbertSp}$ and $\Phi\in\kspositivitymapsb{k}
{\HilbertSp}$. This proves 2).

\noindent $2)\Rightarrow 3)$ This implication is obvious because 
$\kspositivitymapsb{k}{\HilbertSp}\subset\cpositivitymapsb{\HilbertSp}$.

\noindent $3)\Rightarrow 4)$ We know from $3)$ that $\Psi\Phi$ is 
completely positive. As a consequence of Choi's theorem (Proposition \ref{Choithm}),
 $\fnc{J}{\Psi\Phi}=\left(\identitymap\otimes\Psi\Phi\right)
\proj{\psi_+}$ is positive. Thus we have $\Trb{\proj{\psi_+}
\fnc{J}{\Psi\Phi}}\geqslant 0$, which is precisely the statement in $4)$. 

\noindent $4)\Rightarrow 1)$ Note that $\Trb{\proj{\psi_+}\left(\identitymap\otimes\Psi\Phi\right)
\left(\proj{\psi_+}\right)}=\left(\fnc{J}{\mathbbm{1}}|\fnc{J}{\Psi\Phi}\right)$, where $\mathbbm{1}$ is the identity operation and  $\left(.|.\right)$ denotes the Hilbert-Schmidt product in $\lvov$ (cf. Section \ref{sec.GeoOp}). From equality \eqref{Jiso_duality}, we get $\left(\fnc{J}{\mathbbm{1}}|\fnc{J}{\Psi\Phi}\right)=\left(\mathbbm{1}|\Psi\Phi\right)$, which is equal $\left(\Psi^{\ast}|\Phi\right)$. Hence the condition in $4)$ is equivalent to
\begin{equation}\label{equivcond4_1}
 \left(\Psi^{\ast}|\Phi\right)\geqslant 0\,\forall_{\Psi\in\kpositivitymapsb{k}{\HilbertSp}}.
\end{equation}
Using Proposition \ref{prop.takingadjoints} again, 
we see that \eqref{equivcond4_1} is equivalent to
\begin{equation}\label{pointfourduality2}
\left(\Psi|\Phi\right)\geqslant 0\,
\forall_{\Psi\in\kpositivitymapsb{k}{\HilbertSp}},
\end{equation}
Comparing this with the definition \eqref{dualityrefJ-1hvov} 
of the dual cone of $\kpositivitymapsb{k}{\HilbertSp}$ and 
using Proposition \ref{ksposdualkpos}, we obtain
\begin{equation}\label{PhiinkSP}
 \Phi\in\kpositivitymapsb{k}{\HilbertSp}^{\circ}=
\kspositivitymapsb{k}{\HilbertSp},
\end{equation}
which is $1)$.
\end{proof}
\end{theorem}
The following characterization theorem\footnote{and two other ones, cf. \cite{SSZ09}} can be 
proved in practically the same way as Theorem \ref{characterizationsksp}.
\begin{theorem}\label{characterizationskp1}
Let $\Phi\in J^{-1}\left(\hvov\right)$ and $k\in\setN$. The following conditions are equivalent:
\begin{enumerate}[1)]
\item $\Phi\in\kpositivitymapsb{k}{\HilbertSp}$,
\item $\Psi\Phi\in\kspositivitymapsb{k}{\HilbertSp}\,
\forall_{\Psi\in\kspositivitymapsb{k}{\HilbertSp}}$,
\item $\Psi\Phi\in\cpositivitymapsb{\HilbertSp}\,
\forall_{\Psi\in\kspositivitymapsb{k}{\HilbertSp}}$,
\item $\Trb{\proj{\psi_+}\left(\identitymap\otimes
\Psi\Phi\right)\left(\proj{\psi_+}\right)}\geqslant 0\,\,
\forall_{\Psi\in\kspositivitymapsb{k}{\HilbertSp}}$.
\end{enumerate}
\begin{proof}
 It is sufficient to use Proposition \ref{kposdualkspos} instead of \ref{ksposdualkpos} in the last step of the proof of Theorem \ref{characterizationsksp}.
\end{proof}
\end{theorem}
Using Propositions \ref{prop.kPoskBP}, \ref{propkSPkSep}, equation \eqref{Jiso_duality} and the definition \eqref{circmul} of the circled product, one can also formulate the above propositions for subsets of $\hvov$.
\begin{theorem}\label{thm.charkS.1}Let $A\in\hvov$ and $k\in\setN$. The following conditions are equivalent:
\begin{enumerate}[1)]
\item $A\in\ksepopsb{k}{\vov}$,
\item $B\odot A\in\ksepopsb{k}{\vov}\,\forall_{B\in\kbposopsb{k}{\vov}}$,
\item $B\odot A\in\pvov\,\forall_{B\in\kbposopsb{k}{\vov}}$,
\item \textnormal{$\fnc{\Tr}{\fnc{J^{-1}}{B\odot A}}\geqslant 0\,\,\forall_{B\in\kbposopsb{k}{\vov}}$}.
\end{enumerate}
\begin{proof}Let us consider $\Psi=J^{-1}\left(A\right)$ and $\Phi=J^{-1}\left(B\right)$. Conditions $1)-3)$ above are in an obvious one-to-one corresponce with the conditions $1)-3)$ in Theorem \ref{characterizationsksp} (cf. Propositions \ref{prop.kPoskBP}, \ref{propkSPkSep}, eq.\eqref{Jiso_duality} and def.\eqref{circmul}). It remains to be proved that condition $4)$ above corresponds to condition $4)$ in Theorem  \ref{characterizationsksp}. This is also easy to show because $\fnc{\Tr}{\fnc{J^{-1}}{B\odot A}}=\Tr\left(\Psi\Phi\right)=\left(\identitymap|\Psi\Phi\right)$. We have already showed in the proof of Theorem \ref{characterizationsksp} that this expression is equal to $\Trb{\proj{\psi_+}\left(\identitymap\otimes
\Psi\Phi\right)\left(\proj{\psi_+}\right)}$. Thus condition $4)$ above is the same as condition $4)$ in Theorem \ref{characterizationsksp}.
\end{proof}
\end{theorem}
\begin{theorem}\label{thm.charkBP.1}Let $B\in\hvov$ and $k\in\setN$. The following conditions are equivalent:
\begin{enumerate}[1)]
\item $B\in\kbposopsb{k}{\vov}$,
\item $B\odot A\in\ksepopsb{k}{\vov}\,\forall_{A\in\ksepopsb{k}{\vov}}$,
\item $B\odot A\in\pvov\,\forall_{A\in\ksepopsb{k}{\vov}}$,
\item $\fnc{\Tr}{\fnc{J^{-1}}{B\odot A}}\geqslant 0\,\forall_{A\in\ksepopsb{k}{\vov}}$.
\end{enumerate}
\begin{proof}
Follows from Theorem \ref{characterizationskp1} in the same way as Theorem \ref{thm.charkS.1} follows from Theorem \ref{characterizationsksp}.
\end{proof}
\end{theorem}
Note that Theorems \ref{characterizationsksp} and \ref{characterizationskp1} are a broad generalization of a number of relatively well known facts 
about the sets $\positivitymapsb{\HilbertSp}$, $\cpositivitymapsb{\HilbertSp}$ and 
$\spositivitymapsb{\HilbertSp}$,
\begin{eqnarray}
 \label{conditionsSP}
\Phi\in\spositivitymapsb{\HilbertSp}&\Longleftrightarrow&
\Psi\Phi\in\cpositivitymapsb{\HilbertSp}\,
\forall_{\Psi\in\positivitymapsb{\HilbertSp}}\\
\label{conditionsCP}
\Phi\in\cpositivitymapsb{\HilbertSp}&\Longleftrightarrow&
\Psi\Phi\in\cpositivitymapsb{\HilbertSp}\,
\forall_{\Psi\in\cpositivitymapsb{\HilbertSp}}\\
\label{conditionsP}
\Phi\in\positivitymapsb{\HilbertSp}&\Longleftrightarrow&
\Psi\Phi\in\cpositivitymapsb{\HilbertSp}\,
\forall_{\Psi\in\spositivitymapsb{\HilbertSp}}
\end{eqnarray}
(these can be found on page 345 of \cite{ref.Zyczkowski}). Our theorems can also easily be used to prove generalizations of the positive maps criterion by Horodeccy \cite{ref.Horodeccy}. For example,
\begin{corollary}[$k$-positive maps criterion]\label{prop.kposmapscr} Let $\rho\in\hvov$. The operator $\rho$ is $k$-entangled if and only if it satisfies
\begin{equation}\label{kposmapscr1}
 \left(\Psi\otimes\mathbbm{1}\right)\rho\geqslant 0\,\forall_{\Psi\in\kpositivitymapsb{k}{V}}.
\end{equation}
\begin{proof}
 Let us denote $\Phi:=J^{-1}\left(\rho\right)$. Condition \eqref{kposmapscr1} can be rewritten as
\begin{equation}\label{kposmapscr2}
 \left(\Psi\otimes\identitymap\right)J\left(\Phi\right)=\left(\Psi\otimes\identitymap\right)\left(\Phi\otimes\identitymap\right)\proj{\psi_+}=\fnc{J}{\Psi\Phi}\geqslant 0\,\forall_{\Psi\in\kpositivitymapsb{k}{V}}.
\end{equation}
According to the Choi's theorem (Theorem \ref{Choithm}), the above condition is equivalent to
\begin{equation}\label{kposmapscr3}
 \Psi\Phi\in\cpositivitymapsb{V}\,\forall_{\Psi\in\kpositivitymapsb{k}{V}}.
\end{equation}
From Theorem \ref{characterizationsksp}, point $3)$, we get $\Phi=\fnc{J}{\rho}\in\kspositivitymapsb{k}{V}$. According to Proposition \ref{propkSPkSep}, this is the same as $\rho\in\ksepopsb{k}{\vov}$.
\end{proof}
\end{corollary}
For $k=1$, we recover the positive maps separability criterion by Horodeccy (remember that $\ksepopsb{1}{V}$ is the set of separable states). For $k=2$, the above theorem has already appeared, in a little less explicit form, in \cite{Cl05}. 

It is natural to ask whether conditions similar to \eqref{conditionsSP}, \eqref{conditionsCP} or \eqref{conditionsP} hold for $\Phi$ in an arbitrary convex cone $\mathcal{K}\subset J^{-1}\left(\hvov\right)$ (in place of $\spositivitymapsb{V}$, $\cpositivitymapsb{
V}$, $\positivitymapsb{V}$) and $\Psi$ in its dual cone $\mathcal{K}^{\circ}$ (as $\positivitymapsb{V}$, $\cpositivitymapsb{
V}$, $\spositivitymapsb{V}$, resp.). We know they hold for all the cones $\kpositivitymapsb{k}{V}$, $\kspositivitymapsb{k}{V}$ (Theorems \ref{characterizationsksp} and \ref{characterizationskp1}). In the example below, we show that they are not true for general $\mathcal{K}$.
\begin{example}\label{exampleKKdual}
 Consider $\mathcal{K}=\left\{\lambda\identitymap|\lambda\in\left[0;+\infty\right)\right\}$,
where $\identitymap$ refers to identity map acting on $\lv$. Obviously, $\mathcal{K}$ is a closed convex cone in $J^{-1}\left(\hvov\right)$. There exist maps $\Phi\in\mathcal{K}$ and $\Psi\in\mathcal{K}^{\circ}$ s.t.
\begin{equation}\label{phipsicounterex1}
 \Psi\Phi\not\in\cpositivitymapsb{V}.
\end{equation}
\begin{proof}
 From the definitions of the Hilbert-Schmidt product in $\fnc{L}{\lv}$ and the dual cone, $\mathcal{K}^{\circ}=\left\{\Psi\in J^{-1}\left(\hvov\right)|\Tr\Psi\leqslant 0\right\}$. We may assume w.l.o.g. $\Phi=\identitymap$. Thus $\Psi\Phi=\Psi\in\mathcal{K}^{\circ}$ and it can be any element of $\mathcal{K}^{\circ}$. But the defining condition $\Tr\Psi\geqslant 0$ for $\mathcal{K}^{\circ}$ does not imply $\Psi\in\cpositivitymapsb{V}$. For example, the transposition map $t:\pprj{\alpha}{\beta}\mapsto\pprj{\beta}{\alpha}$ has $\Tr t=\sum_{\alpha,\beta=1}^d\left(\pprj{\alpha}{\beta}|\pprj{\beta}{\alpha}\right)=\sum_{\alpha=1}^d\left(\pprj{\alpha}{\alpha}|\pprj{\alpha}{\alpha}\right)=d>0$, but it is well-known that $t$ is not a completely positive map (cf. e.g. \cite{ref.Zyczkowski}).
\end{proof}
\end{example}
A correct way to generalize Theorems \ref{characterizationsksp} and \ref{characterizationskp1} has recently been found by St\o rmer \cite{St08,SSZ09,St09symmetric}. Let us briefly describe it. For an arbitrary $C^{\ast}$-algebra $A$ and a Hilbert space $\mathcal{H}$, one considers bounded linear
 maps $\Phi$ of $A$ into $\fnc{B}{\mathcal{H}}$ -- the space of bounded linear operators on $\mathcal{H}$. For any such $\Phi$, there exists a corresponding linear functional
 $\tilde\Phi$ on $A\otimes\fnc{B}{\mathcal{H}}$ given by
\begin{equation}\label{phitilde}
 \tilde\Phi\left(a\otimes b\right)=\Trb{\Phi\left(a\right)b^{\transpos}},\,a\in
 A,\,b\in\fnc{B}{\mathcal{H}},
\end{equation}
where $\Tra{}$ is the usual trace on $\fnc{B}{\mathcal{H}}$ and $\transpos$ the
 transpose. The correspondence $\Phi\leftrightarrow\tilde\Phi$ is an analogue of $J$ in the more general setting described above. Let ($\cpositivitymapsb{\mathcal{H}}$) $\positivitymapsb{\mathcal{H}}$ denote the set of (completely) positive maps of $B\left(\mathcal{H}\right)$ into itself. We say that a nonzero cone
 $\mathcal K$ in $\positivitymapsb{\mathcal{H}}$ is a \textit{mapping cone} if
 $\Phi\in\mathcal{K}$ implies $\Psi\Phi\Upsilon\in\mathcal{K}$ for all
 $\Psi,\Upsilon\in\cpositivitymapsb{\mathcal{H}}$. It turns out (cf. \cite{SSZ09}) that all the cones $\kpositivitymapsb{k}{V}$, $\kspositivitymapsb{k}{V}$ ($k=1,\ldots,d$) discussed above are examples of mapping cones.
\begin{proposition}Take $\mathcal{H}=V$ with $V$ as in the earlier parts of this section. Then $B\left(\mathcal{H}\right)=\lv$. Let $A$ be also equal to $\lv$. All the cones $\kpositivitymapsb{k}{V}$ and $\kspositivitymapsb{k}{V}$ are mapping cones.
\qed
\end{proposition}
For an arbitrary mapping cone $\mathcal{K}\subset\positivitymapsb{\mathcal{H}}$, one defines
\begin{equation}\label{defPAK}
 \PAK{A}{\mathcal{K}}:=\left\{x\in A\otimes\fnc{B}{\mathcal{H}}|\,x=\hconj{x},\identitymap\otimes\Psi\left(x\right)
\geqslant 0\,\forall_{\Psi\in\mathcal{K}}\right\},
\end{equation}
where $\identitymap$ denotes the identity map on $\fnc{B}{\mathcal{H}}$.
$\PAK{A}{\mathcal{K}}$ is a proper closed cone in $A\otimes\fnc{B}{\mathcal{H}}$. A given map $\Phi$ of $A$ into $B\left(\mathcal{H}\right)$ is called \textit{$\mathcal K$-positive} if $\tilde\Phi$ is positive on
$\PAK{A}{\mathcal{K}}$. Let us denote the set of $\mathcal K$-positive maps of $A$ into $\fnc{B}{\mathcal{H}}$ with $\Kapositivemapsb{\mathcal K}{\mathcal{H}}$. One can prove \cite{SSZ09} that
\begin{theorem}\label{thm42Erling}Take $\mathcal{H}=V$, $B\left(\mathcal{H}\right)=\lv$ and $A=\lv$. We have
\begin{equation}\label{PKforSPkBP}
 \kspositivitymapsb{k}{\HilbertSp}=\Kapositivemapsb{\kspositivitymapsb{k}{\HilbertSp}}{\HilbertSp}\quad\textnormal{and}\quad \kpositivitymapsb{k}{\HilbertSp}=\Kapositivemapsb{\kpositivitymapsb{k}{\HilbertSp}}{\HilbertSp}
\end{equation}
for arbitrary $k=1,2,\ldots,d$.\qed
\end{theorem}

In the language of mapping cones, condition $3)$ in Theorem \ref{characterizationsksp} or in Theorem \ref{characterizationskp1} corresponds to the fact that $\kspositivitymapsb{k}{V}$ or $\kpositivitymapsb{k}{V}$ (resp.) fulfil the properties \eqref{PKforSPkBP}. In general, a cone $\mathcal{K}\subset\positivitymapsb{\mathcal{H}}$ can be characterized in a similar way as in point $3)$ of Theorems \ref{characterizationsksp}, \ref{characterizationskp1} if it fulfils $\mathcal{K}=\Kapositivemapsb{\mathcal{K}}{\mathcal{H}}$. In the paper \cite{St09symmetric}, it is proved that $\mathcal{K}=\Kapositivemapsb{\mathcal{K}}{\mathcal{H}}$ holds for arbitrary {\it symmetric mapping cones}, i.e. cones $\mathcal{K}$ s.t. $\Phi\in\mathcal{K}\Rightarrow \hconj{\Phi}\in\mathcal{K}$ and $\Phi\in\mathcal{K}\Rightarrow t\Phi t\in\mathcal{K}$, where $t$ denotes the transposition map and $\Phi^{\ast}$ is the adjoint of $\Phi$ w.r.t. the Hilbert-Schmidt product in $B\left(\mathcal{H}\right)$. All the cones $\kpositivitymapsb{k}{V}$, $\kspositivitymapsb{k}{V}$ ($k=1,\ldots,d$) are examples of symmetric mapping cones.


\section{Positivity conditions on entanglement witnesses}\label{sec.WitPosmaps}
The positive maps criterion by Horodeccy \cite{ref.Horodeccy} (i.e.\,our Corollary \ref{prop.kposmapscr} with $k=1$) allows to detect entanglement of a given state $\rho\in\mvov$ by checking that $\left(\Phi\otimes\mathbbm{1}\right)\rho$ is not a positive operator for some positive map $\Phi$. This gives an obvious way to obtain a separability criterion from a positive map. Note, however, that only the maps that are not completely positive are suitable for that purpose. Among the best known ones are the transposition map, leading to the PPT criterion \cite{ref.Peres,ref.Horodeccy}, and the map $\Lambda:\xi\mapsto\Tr\xi\,\mathbbm{1}-\xi$, giving rise to so-called reducion criterion \cite{HH99}.

As a consequence of Propositions \ref{Jamiolkowskithm} and \ref{prop.kSepdualkBP} (for $k=1$), $J$-transforms of positive maps $\Phi$ such that $\left(\fnc{J}{\Phi}|\rho\right)<0$ for some $\rho\in\mvov$  can be used to detect entanglement of $\rho$. According to \ref{prop.kSepdualkBP}, $\left(\fnc{J}{\Phi}|\rho\right)<0$ implies nonseparability of $\rho$. Therefore operators $A=\fnc{J}{\Phi}\in\bposopsb{V}$ with the property $\left(A|\rho\right)<0$ for some $\rho$ are called {\it entanglement witnesses}. Again, as a consequence of Choi's theorem (Theorem \ref{Choithm}), only non-completely positive maps $\Phi$ yield entanglement witnesses. The value of $\left(A|\rho\right)$ can be relatively easily measured in experiments (cf.\,e.g.\cite{GHB+03}), which is the reason why the entanglement witness approach is usually preferred to the more efficient positive maps criterion mentioned earlier. In any case, characterization of {\it positive maps in $\fnc{L}{\lv}$ that are not completely positive} is crucial for understanding the structure of the set of separable states and thus the nature of entanglement. Equivalently, one may consider operators in $\lvov$ that are {\it block positive but not positive}. This follows from the theorems by Jamio\l kowski (Proposition \ref{Jamiolkowskithm}) and Choi (Theorem \ref{Choithm}). Since positivity of an operator can be checked by elementary methods (cf. equation \eqref{minorspos} below), the only remaining problem consists in checking the block positivity condition \eqref{bposdef}. This is by far not a trivial task to do and we shall only give partial results concerning it. In general, the problem remains open. One of our results  does even say that block positivity cannot be checked in an affirmative way using a wide class of criterions that we introduce in Section \ref{sec.neccesaryimsufficient}. 

In this part of the paper, we always work with tensor products of the form $V\otimes V$. We decided to do so for the sake of simpler discussion. Our results can easily be formulated for tensor products $U\otimes V$ with $U\neq V$, just as it was done in \cite{SZ09}.

\subsection{Block positivity condition and positive polynomials}\label{sec.BPPospolys}Let us remind the reader that the block positivity condition \eqref{bposdef} for an operator $A\in\lvov$ reads
\begin{equation}\label{fml.defBP.2}
 \hprod{\upsilon\otimes u}{\fnc{A}{\upsilon\otimes u}}\geqslant 0\,\forall_{\upsilon,u\in V}.
\end{equation}
We have already mentioned in Section \ref{sec.kpospsposdual} that \eqref{fml.defBP.2} implies  Hermiticity of $A$ (for a proof, cf. \cite{pracamag}). In index notation, for some choice of orthonormal basis \SEQ{\left|\alpha\right>}{\alpha=1}{d} of $V$, condition \fml{defBP.2} reads
\begin{equation}\label{fml.defBP.3}
\left(\upsilon^{\alpha}\right)^{\ast}\left(u^{\beta}\right)^{\ast}A_{\alpha\beta,\gamma\delta}\upsilon^{\gamma}u^{\delta}\geqslant 0\,\forall_{\seq{\upsilon^{\alpha}}{\alpha=1}{d},\seq{u^{\beta}}{\beta=1}{d}\subset\mathbbm{C}},
\end{equation}
where $A_{\alpha\beta,\gamma\delta}$ are matrix elements of $A$ (that is, $\fnc{A}{\left|\gamma\right>\left|\delta\right>}=\sum_{\alpha,\beta=1}^{d} A_{\alpha\beta,\gamma\delta}\left|\alpha\right>\left|\beta\right>$).

To rewrite condition \eqref{fml.defBP.3} in a more convenient way, we define a family of Hermitian operators $\check{A}_{\upsilon}$ and $\skew6\hat{A}_u$ on $V$ with matrix elements $\left(\check{A}_{\upsilon}\right)_{\beta\delta}:=\left(\upsilon^{\alpha}\right)^{\ast}A_{\alpha\beta,\gamma\delta}\upsilon^{\gamma}$ and $\left(\skew6\hat{A}_u\right)_{\alpha\gamma}:=\left(u^{\beta}\right)^{\ast}A_{\alpha\beta,\gamma\delta}u^{\delta}$. The block positivity property \fml{defBP.3} means that the operators $\check{A}_{\upsilon}$ and $\skew6\hat{A}_u$ are positive for arbitrary $\upsilon$ and $u$. We can rewrite \fml{defBP.3} either as
\begin{equation}\label{fml.defBP.4}
 \check{A}_{\upsilon}\geqslant 0\,\forall_{\upsilon\in V}\quad\textrm{or as}\quad\skew6\hat{A}_u\geqslant 0\,\forall_{u\in V}.
\end{equation}
 
The inequality signs in \fml{defBP.4} 
refer to positivity of operators. Although the above two conditions look better than \fml{defBP.3}, they are actually not simple at all because of the different meaning of inequality signs. Nevertheless, we prefer to use the expressions in \fml{defBP.4}, 
 which involve a reduced number of free parameters. The operators $\check{A}_{\upsilon}$ and $\skew6\hat{A}_{u}$ may be called {\it blocks} of $A$, which explains the name of block positivity to some extent.

According to \fml{defBP.4}, a Hermitian operator $A$ on $\vov$ is block positive if and only if all its blocks are positive operators, though we are allowed to check it for $\check A_{\upsilon}$'s or for $\skew6\hat{A}_u$'s alone. In the following, we shall concentrate on the right hand side of \eqref{fml.defBP.4}. That is, we will be discussing the condition ${\skew6\hat{A}_u}\geqslant 0$.

Let us denote with $[\skew6\hat{A}_u]$ the matrix of $\skew6\hat{A}_u$ in the chosen basis of $V$. Positivity of $\skew6\hat{A}_u$ for all $u$ is equivalent to the following set of inequalities,
\begin{eqnarray}
\label{minorspos}
 W_l\left(u\right):=\sum_{1\leqslant i_1<i_2<\ldots<i_l\leqslant N_2}\Delta_{i_1i_2\ldots i_l}\left(\skew6\hat{A}_u\right)
\geqslant 0\,\forall_{u\in V}\forall_{l=1\ldots N_2},
\end{eqnarray}
where $\Delta_{i_1i_2\ldots i_l}(\skew6\hat{A}_u)$ is the minor of $[\skew6\hat{A}_u]$ 
involving only columns and rows with the numbers $i_1,\ldots,i_l$. It follows from the discussion in \cite{Jamiolkowski1} 
that\footnote{an explicit proof can be found in \cite{pracamag}} the functions $W_l$ are homogeneous real polynomials of an even degree in the variables 
$\seq{\Repart{u^{\alpha}}}{\alpha=1}{N_1}$, $\seq{\Impart{u^{\gamma}}}{\gamma=1}{N_1}$. 
Thus \eqref{minorspos} is a set of positivity conditions for real homogeneous polynomials of an even degree. 
If we can check these conditions, we can answer the question whether a given operator $A$ is block positive. However, no simple method for checking positivity of real homogenous polynomials seems to be known. It is in principle possible to eliminate quantifiers from formulas like $\forall_{\left\{x_1,\ldots,x_{n-1}\right\}\subset\setR}\sum_{i_1,i_2,\ldots,i_n}C_{i_1i_2\ldots i_n}x_1^{i_1}\ldots x^{i_n}_n\geqslant 0$, e.g. using Gr\"obner bases,
 but the outcome usually involves 
\emph{zeros of univariate polynomials of a very high degree}. These cannot in general be expressed in terms of the coefficients $C_{i_1i_2\ldots i_n}$. Also numerical solutions to these conditions may often prove useless in practice, because a very high precision is needed to make the results of the whole procedure reliable.

Fortunately, there exist situations where the approach suggested above leads to explicit block positivity conditions. To demonstrate this, let $a$, $b$ and $c$ be arbitrary complex numbers and consider the following family of matrices,
\begin{equation}\label{abcfamilydef}
 \left[A\left(a,b,c\right)\right]=\matrfour{A}=\left[\begin{array}{cccc}
               	\frac{1}{2}&a&0&0\\
		\bar a&\frac{1}{2}&b&0\\
		0&\bar b&\frac{1}{2}&c\\
		0&0&\bar c&\frac{1}{2}\\
              \end{array}\right],
\end{equation}
corresponding to a family of Hermitian operators $A\left(a,b,c\right)$ on $\setC^2\otimes\setC^2$. In order to test condition \eqref{fml.defBP.3} using the method suggested above, observe that
\begin{equation}
 \label{blocksabc}\left[\skew6\hat{A}_u\left(a,b,c\right)\right]=\left[
  \begin{array}{cc}
    \frac{1}{2}\left(\left|u_1\right|^2+\left|u_2\right|^2\right)&a\left|u_1\right|^2+c\left|u_2\right|^2+\bar bu_1\bar u_2\\
    \bar a\left|u_1\right|^2+\bar c\left|u_2\right|^2+b\bar u_1 u_2&\frac{1}{2}\left(\left|u_1\right|^2+\left|u_2\right|^2\right)\\
  \end{array}
 \right].
\end{equation}
Obviously, $\skew6\hat{A}_u\left(a,b,c\right)$ is a positive operator for all $u\in\setC^2$ if and only if its determinant satisfies $\det \skew6\hat{A}_u\left(a,b,c\right)\geqslant 0\,\forall_{u\in\setC^2}$. That is,
\begin{equation}\label{conditionabc1}
 \left(\frac{1}{2}\left(\left|u_1\right|^2+\left|u_2\right|^2\right)\right)^2-\left|a\left|u_1\right|^2+c\left|u_2\right|^2+
\bar bu_1\bar u_2\right|^2\geqslant 0\,\forall_{u_1,u_2\in\setC}.
\end{equation}
In \cite{SZ09}, we show that \eqref{conditionabc1} is equivalent to 
\begin{equation}
  \label{conditionabc4}
   1-\left|\alpha+\gamma\cos\varphi\right|-\left|b\right|\sin\varphi\geqslant 0\,\forall_{\varphi\in\setR},
 \end{equation}
where $\alpha:=a+c$ and $\gamma:=a-c$. Condition \eqref{conditionabc4} can be easily solved in the two following situations: 
\begin{itemize}
\item[a)] $\Repart{\alpha\bar\gamma}=0 \ \Longleftrightarrow \  \left|a\right|=\left|c\right|$
\item[b)] $\Repart{\alpha\bar\gamma}=\pm\left|\alpha\right|\left|\gamma\right| \  \Longleftrightarrow \ a=rc,\, r\in\setR$
\end{itemize}
In the case a), condition \eqref{conditionabc4} simplifies to
\begin{equation}
  \label{conditionsita}
 1-\sqrt{\left|\alpha\right|^2+\left|\gamma\right|^2\cos^2\varphi}-\left|b\right|\sin\varphi\geqslant 0\,\forall_{\varphi\in\setR}.
\end{equation}
We observe that $\left|\alpha\right|^2+\left|\gamma\right|^2\leqslant 1$ must hold in order that \eqref{conditionsita} be true.
 Keeping this in mind, we can rewrite \eqref{conditionsita} as
\begin{equation}\label{conditionsita2}
 \left|\frac{b}{\gamma}\right|^2\lambda^2-\lambda+\left(1-\left|\frac{b}{\gamma}\right|^2\left(\left|\alpha\right|^2+
\left|\gamma\right|^2\right)\right)\geqslant 0\,\forall_{\lambda\in\left[\left|\alpha\right|,\sqrt{\left|\alpha\right|^2
+\left|\gamma\right|^2}\right]}.
\end{equation}
where we substituted $\sqrt{\left|\alpha\right|^2+\left|\gamma\right|^2\cos^2\varphi}\rightarrow\lambda$. 
As a positivity condition for a quadratic function, \eqref{conditionsita2} can be easily solved explicitly.
 Together with the condition on $\left|\alpha\right|^2+\left|\gamma\right|^2$, we obtain
\begin{equation}
\label{conditionsitaexpl}
 \left|\alpha\right|^2+\left|\gamma\right|^2\leqslant 1\land\left|\alpha\right|+
 \left|b\right|^2\leqslant 1\land\left\{2\left|b\right|^2\left|\alpha\right|\leqslant\left|\gamma\right|^2\lor
  2\left|b\right|^2\sqrt{\left|\alpha\right|^2+\left|\gamma\right|^2}\geqslant\left|\gamma\right|^2\right\}.
\end{equation}
In the case b), it is even simpler to get the conditions on $\alpha$, $\gamma$ and $b$ equivalent to \eqref{conditionabc4}. We have $\left|\alpha+\gamma\cos\varphi\right|\leqslant\left|\alpha\right|+\left|\gamma\right|\left|\cos\varphi\right|$. Either for $\varphi$ or for $\varphi\rightarrow\pi-\varphi$, we obtain $\left|\alpha+\gamma\cos\varphi\right|=\left|\alpha\right|+\left|\gamma\right|\left|\cos\varphi\right|$ and $\sin\varphi$ is not changed by the substitution $\varphi\rightarrow\pi-\varphi$. Hence we can rewrite \eqref{conditionabc4} as
\begin{equation}\label{casebconditions1}
 1-\left|\alpha\right|-\left|\gamma\right|\left|\cos\varphi\right|-\left|b\right|\sin\varphi\geqslant 0\,\forall_{\varphi\in\setR}.
\end{equation}
This is equivalent to $\left(1-\left|\alpha\right|-\left|\gamma\right|\cos\varphi-\left|b\right|\sin\varphi\right)\geqslant 0\forall_{\varphi\in\setR}$, which is easy to solve explicitly in terms of $\alpha$, $\gamma$ and $b$. We get
\begin{equation}\label{casebconditions2}
 1-\left|\alpha\right|-\sqrt{\left|\gamma\right|^2+\left|b\right|^2}\geqslant 0.
\end{equation}
For general $a,b,c\in\setC$, the biggest obstacle in solving \eqref{conditionabc4}
is a positivity condition on $\left[-1;1\right]$ for a polynomial of degree 4. This problem can in principle be solved explicitly in terms of the coefficients of the polynomial, but we have not been able to bring the solution to a readable form. In any case, given some particular values of $a$, $b$ and $c$, one can easily check condition \eqref{conditionabc4} using, for example, Sturm sequences \cite{Marten}. More details can be found on page 6 of \cite{pracamag}. We should also remark that the results presented in this section are not a simple consequence of the St\o rmer-Woronowicz theorem \cite{ref.Stormer,ref.Woronowicz} about decomposability of maps of $2\times 2$ matrices, even though $A\left(a,b,c\right)$ was acting on a $2\times 2$-dimensional space. 

\subsection{A family of necessary but insufficient criterions}\label{sec.neccesaryimsufficient}
The positivity conditions in \fml{defBP.4} can be used in an obvious way to produce various criterions for block positivity.  For this, we take two finite sets \SEQ{\upsilon_i}{i=1}{n} and \SEQ{u_j}{j=1}{m} of vectors in $V$ and consider the following set of conditions,
\begin{equation}\label{fml.BPcrit.1}
\seq{\forall_{i=1\ldots n}\check{A}_{\upsilon_i}\geqslant 0}{}{}{}\land\seq{\forall_{j=1\ldots m}\skew6\hat{A}_{u_j}\geqslant 0}{}{}.
\end{equation}
As a consequence of \fml{defBP.4}, \eqref{fml.BPcrit.1} is a necessary criterion for block positivity for any choice of the vectors $\upsilon_i$ and $u_j$. Positivity of $\check{A}_{\upsilon_i}$ and $\skew6\hat{A}_{u_j}$ can be checked by elementary methods (cf. equation \eqref{minorspos}), which looks promissing. Nevertheless, criterions of the form \fml{BPcrit.1} can never be sufficient, as we show in the following proposition. 
\begin{proposition}\label{prop.insufficient}Let $\seq{\upsilon_i}{i=1}{n},\seq{u_j}{j=1}{m}\subset V$ be two finite sets of vectors in $V$. There exists an operator $A\in\hvov$ which is not block positive, but it fulfils the conditions \eqref{fml.BPcrit.1}.
\begin{proof}
To to prove our assertion, we choose two vectors $\upsilon_0,u_0\in V$ in such a way that $\upsilon_0$ is not proportional to any of the vectors $\upsilon_j$ ($j=1\ldots n$) and $u_0$ is not proportional to any of the vectors $u_j$ ($j=1\ldots m$). We may assume w.l.o.g. that all $\upsilon_i$'s and $u_j$'s (including $\upsilon_0$ and $u_0$) are of unit norm. From the definition of $u_0$ and $\upsilon_0$ it then follows that the numbers $\nu:=\max_{i=1\ldots n}\seq{\left|\hprod{\upsilon_i}{\upsilon_0}\right|^2}{}{}$ and $\mu:=\max_{i=1\ldots m}\seq{\left|\hprod{u_i}{u_0}\right|^2}{}{}$ are lower than 1. Define the following Hermitian operators on $V$,
\begin{equation}\label{fml.opN.1}
 N:=\frac{1+\nu}{1-\nu}\left(\id-\prj{\upsilon_0}\right)-\prj{\upsilon_0}
\end{equation}
and
\begin{equation}\label{fml.opM.1}
 M:=\frac{1+\mu}{1-\mu}\left(\id-\prj{u_0}\right)-\prj{u_0}.
\end{equation}
It is easy to see that neither $N$ nor $M$ is positive. We have $\hprod{\upsilon_0}{\fnc{N}{\upsilon_0}}=-1$ and $\hprod{u_0}{\fnc{M}{u_0}}=-1$ and these are the minimum values of $\hprod{\upsilon}{\fnc{N}{\upsilon}}$ and $\hprod{u}{\fnc{M}{u}}$ over all normalized vectors $\upsilon$ and $u$, resp. Let us take
\begin{equation}\label{fml.opA.1}
 A=N\otimes\id+\id\otimes M.
\end{equation}
We have $\hprodA{A}{\upsilon_0\otimes u_0}=\hprodA{N}{\upsilon_0}\hsqrt{u_0}+\hsqrt{\upsilon_0}\hprodA{M}{u_0}=-2$ and thus $A$ is not block positive. On the other hand, we will show that conditions \fml{BPcrit.1} hold. To do that, we first observe that $\check{A}_{\upsilon_i}=\hprodA{N}{\upsilon_i}\id+M$ and $\skew6\hat{A}_{u_j}=\hprodA{M}{u_j}\id+N$. Explicit expressions for $\hprodA{N}{\upsilon_i}$ and $\hprodA{M}{u_j}$ read
\begin{equation}\label{fml.opN.2}
 \hprodA{N}{\upsilon_i}=\frac{1+\nu}{1-\nu}-\left|\hprod{\upsilon_i}{\upsilon_0}\right|^2\frac{2}{1-\nu}
\end{equation}
and
\begin{equation}\label{fml.opM.2}
 \hprodA{M}{u_j}=\frac{1+\mu}{1-\mu}-\left|\hprod{u_j}{u_0}\right|^2\frac{2}{1-\mu}.
\end{equation}
Because of the definition of $\nu$, $\hprodA{N}{\upsilon_i}\geqslant\frac{1+\nu}{1-\nu}-\nu\frac{2}{1-\nu}=1$. In a very similar way, $\hprodA{M}{u_j}\geqslant 1$. To prove our assertion, we calculate $\hprodA{\skew6\hat{A}_{u_j}}{\upsilon}$ and $\hprodA{\check{A}_{\upsilon_i}}{u}$ for arbitrary normalized vectors $\upsilon,u\in V$. We get
\begin{equation}\label{fml.opMN.1}
 \hprodA{\skew6\hat{A}_{u_j}}{\upsilon}=\hprodA{M}{u_j}+\hprodA{N}{\upsilon}
\end{equation}
and
\begin{equation}\label{fml.opNM.1}
 \hprodA{\check{A}_{\upsilon_i}}{u}=\hprodA{N}{\upsilon_i}+\hprodA{M}{u}.
\end{equation}
As we already mentioned above, $\hprodA{N}{\upsilon}\ge\hprodA{N}{\upsilon_0}=-1$ and $\hprodA{M}{u}\geqslant\hprodA{M}{u_0}=-1$. We also know that $\hprodA{N}{\upsilon_i}\geqslant 1$ and $\hprodA{M}{u_j}\geqslant 1$, which leads us to $\hprodA{M}{u_j}+\hprodA{N}{\upsilon}\geqslant 0$ and $\hprodA{N}{\upsilon_i}+\hprodA{M}{u}\geqslant 0$. Referring back to \fml{opMN.1} and \fml{opNM.1}, this is precisely
\begin{equation}\label{fml.BPcrit.2}
 \hprodA{\skew6\hat{A}_{u_j}}{\upsilon}\geqslant 0\land\hprodA{\check{A}_{\upsilon_i}}{u}\geqslant 0
\end{equation}
for all $\upsilon,u\in V$ of unit norm. But the normalization assumption is superfluous in \eqref{fml.BPcrit.2} and we see that \eqref{fml.BPcrit.2} is just equivalent to \eqref{fml.BPcrit.1}. Hence \fml{BPcrit.1} is never a sufficient block positivity criterion.
\end{proof}
\end{proposition}
In particular, our result implies to any block positivity criterion that follows from a finite number of conditions of the type \fml{BPcrit.1}. To give an example, let us mention a family of criterions recently proved by Sommers \cite{Sommers},
\begin{multline}\label{HJSommers_criterion}
 \left|\innerpr{\alpha\otimes\mu}{A\left(\beta\otimes\nu\right)}\right|^2\leqslant\frac{1}{2}\bigl(\innerpr{\alpha\otimes\mu}{A\left(\alpha\otimes\mu\right)}\innerpr{\beta\otimes\nu}{A\left(\beta\otimes\nu\right)}\bigr.\\\bigl.+\innerpr{\alpha\otimes\nu}{A\left(\alpha\otimes\nu\right)}\innerpr{\beta\otimes\mu}{A\left(\beta\otimes\mu\right)}\bigr),
\end{multline}
where $\left|\alpha\right>$, $\left|\beta\right>$, $\left|\mu\right>$ and $\left|\nu\right>$ are arbitrary elements of an orthonormal basis $\left\{\left|\alpha\right>\right\}_{\alpha=1}^d$ of $V$. Even though \eqref{HJSommers_criterion} is not of the form \eqref{fml.BPcrit.1}, it can be deduced from a criterion of that type\footnote{we leave the details to the interested reader}. It follows from Proposition \ref{prop.insufficient} that \eqref{fml.BPcrit.1} is not sufficient for checking block positivity. More surprisingly, Proposition \ref{prop.insufficient} tells us that a finite set of conditions like \eqref{HJSommers_criterion} can never yield a sufficient block positivity criterion.


\subsection{Block positivity over the reals. Sums of squares}\label{sec.BPoverreals}
It is natural to ask about a simplified version of the block positivity condition \eqref{fml.defBP.2} where the underlying field $\setC$ is substituted with $\setR$. To formulate the problem, let $X$ be a finite-dimensional vector space of dimension $d$ over $\setR$ with a symmetric inner product $\left(.\right)\cdot\left(.\right)$. We call an operator $A\in\fnc{L}{X\otimes X}$ {\it block positive over $\setR$} if it fulfils
\begin{equation}
  \label{blockposRdef}
  \left(x\otimes y\right)\cdot A\left(x\otimes y\right)\geqslant 0\hskip 3 mm\forall_{x,y\in X},
\end{equation}
where ``$\cdot$'' denotes the symmetric inner product inherited  by $X\otimes X$ from $X$.

Condition \eqref{blockposRdef} does not imply symmetry of $A$, but we may always assume that $A$ is symmetric because the antisymmetric part of $A$ vanishes in \eqref{blockposRdef}. 
 With this assumption, $\left(X\otimes X\right)^2\ni\left(w_1,w_2\right)\mapsto w_1\cdot A\left(w_2\right)\in\setR$ is a symmetric bilinear 
form on $X\otimes X$. We can rewrite \eqref{blockposRdef} using index notation,
\begin{equation}
\label{blockposdefRind}
A_{ab,cd}x^a y^b x^c y^d\geqslant 0\,\forall_{\seq{x^a}{a=1}{d},\seq{y^b}{b=1}{d}\subset\setR},
\end{equation}
where $x^a$ and $y^b$ are coordinates of $x$ and $y$ in some orthonormal basis \SEQ{e_a}{a=1}{d} of $X$, and $A_{ab,cd}$ denote the matrix elements of $A$ w.r.t. $\seq{e_a\otimes e_b}{a,b=1}{d}$.

Obviously, \eqref{blockposdefRind} is a positivity condition for a real multivariate polynomial of degree $4$. One possible reason for \eqref{blockposdefRind} may be that $A_{ab,cd}x^a y^b x^c y^d$ is a \textit{sum of squares}
   (SOS) of a family of polynomials $P_i$. In such case, it must satisfy
\begin{equation}
  \label{SOS}
  A_{ab,cd}x^a y^b x^c y^d \ =\ \sum_{i=1}^nP_i^2 \ = \  \sum_{i=1}^n\left(B^i_{ab}x^ay^b\right)^2,
\end{equation}
with real coefficients $B^i_{ab}$ and with the index $i$ running from $1$ to a finite $n$. The last equality follows because the polynomials $P_i$ must be homogeneous, of degree $2$ and they cannot have terms of the form $x^ax^b$, 
neither of the form $y^ay^b$, since there are no terms $\left(x^ax^b\right)^2$ nor $\left(y^ay^b\right)^2$ in the sum $A_{ab,cd}x^a y^b x^c y^d$.

Note that the last expression in \eqref{SOS} equals $\left(x\otimes y\right)\cdot B\left(x\otimes y\right)$ for a positive definite operator $B$ with matrix elements $\sum_{i=1}^nB^i_{ab}B^i_{cd}$. It is therefore tempting to say that \eqref{SOS} implies positive definiteness 
of $A$, but \emph{this is not true}. In fact, the expression $A_{ab,cd}x^a y^b x^c y^d$ allows for an additional {\it partial transpose symmetry} in $A$, which needs to be taken into account. We have
\begin{proposition}\label{propSOS}
 Let $A$ be an operator on $X\otimes X$ \textit{symmetric with respect to partial transpose}, i.e. $\left(\id\otimes t\right)A=A$ where $t$ denotes the transposition map in $L\left(X\right)$. Denote with $A_{ab,cd}$ the matrix elements of $A_{ab,cd}$ w.r.t. an orthonormal product basis of $X\otimes X$. The polynomial $A_{ab,cd}x^a y^b x^c y^d$ is a SOS if and only if
\begin{equation}\label{fml.sosiff}
 A=B+\left(\identitymap\otimes t\right)B
\end{equation}
for a positive operator $B$ on $V\otimes V$. This is equivalent to $A$ being decomposable.
\begin{proof}
 A detailed proof has been included in \cite{SZ09}. Let us only sketch the main points here. For any $A$ that satisfies \eqref{SOS}, we can define a corresponding operator $\tilde A$ with matrix elements
\begin{equation}
 \label{Atildematr}
 \tilde A_{ab,cd}=\frac{1}{2}\left(\sum_iB^i_{ab}B^i_{cd}+B^i_{ad}B^i_{cb}\right).
\end{equation}
It is easy to see that $\left(x\otimes y\right)\cdot\tilde A\left(x\otimes y\right)=\left(x\otimes y\right)\cdot
 A\left(x\otimes y\right)$ for all $x,y\in X$. In \cite[Appendix A]{SZ09} we show that this property together 
with with the symmetry $\left(\id\otimes t\right)A=A$ imply $\tilde A=A$. But $\tilde A$ is of the form \eqref{fml.sosiff} with $B$ positive.
\end{proof}
\end{proposition} 
Since every operator $A\in\fnc{L}{X\otimes X}$ has a symmetrization $\tilde A=\frac{1}{2}\left(A+\left(\identitymap\otimes t\right)A\right)$ that satisfies both $\left(\id\otimes t\right)\tilde A=\tilde A$ and $\left(x\otimes y\right)\cdot\tilde A\left(x\otimes y\right)=\left(x\otimes y\right)\cdot
 A\left(x\otimes y\right)\,\forall_{x,y\in X}$, Proposition \eqref{propSOS} is the maximum we can tell about $A$, given that $A_{ab,cd}x^a y^b x^c y^d$ is SOS. The relation of sums of squares to condition \eqref{fml.defBP.2} for entanglement witnesses is even less clear, though it is known that Choi used facts concerning SOS to give the first example of an indecomposable positive map \cite{Choi75}.

\section{Length of separable states}\label{sec.cardinality}
It follows from the Carath\'eodory's theorem on convex sets in $\setR^n$ \cite{ref.Rockafellar} that any separable state on $U\otimes V$ can be written as a sum of no more than $d^2h^2+1$ tensor products of positive operators, $d$ and $h$ being the dimension of $V$ and $U$, respectively. In other words, for any state $\rho$ of the form \eqref{sepstates} there exists a $l\leqslant d^2h^2+1$ such that
\begin{equation}\label{sepstatesfinite}
 \rho=\sum_{i=1}^l p_i\rho_i\otimes\sigma_i,
\end{equation}
for some positive operators $\rho_i\in\muu$, $\sigma_i\in\mv$ and positive $p_i$ such that $\sum_ip_i=1$. We call the number $l$ {\it length} of $\rho$, in contrast to the related concept of cardinality, cf. \cite{STV98}. It is desirable to determine either length or cardinality of a separable state $\rho$ as a way to quantify the amount of classical communication between subsystems needed to create $\rho$. The knowledge of length or cardinality\footnote{note that some authors \cite{AS09,L00} use the word 'length' when they refer to cardinality} may also help in finding all possible decompositions of $\rho$ into pure product states (cf. \cite{AS09}) or decompositions of the form \eqref{sepstatesfinite}.  At first glance, the length of a separable state $\rho$ seems very similar to its {\it Schmidt rank}, which is the minimal number $r$ such that
\begin{equation}\label{Schmidtdec}
 \rho=\sum_{i=1}^rA_i\otimes B_i
\end{equation}
for some $A_i\in\lu$ and $B_i\in\lv$ Hermitian. Note, however, that the operators $A_i$ and $B_i$ in \eqref{Schmidtdec} do not have to be positive. Thus $l$ does not have to be equal to $r$ in general. Obviously, $l\leqslant r$. It turns out that $l\leqslant 3$ implies $r=l$, whereas for $l=4$, we give an example of a state with Schmidt rank $3$, thus smaller than $l$. 

Determining the length of a given $\rho$ is an open problem. In the present section, we make first steps toward a characterization of separable states according to their length by relating it to the Schmidt rank. We also show that in the case $U=V$, the number of terms in a decomposition of the form \eqref{sepstatesfinite} does not increase under the circled product defined in Section \ref{sec.Entanglement}. This gives an additional motivation for studying the length of states. It also points to the circled multiplication as a natural product in $V\otimes V$ when entanglement properties of states are considered.

\subsection{Length and the circled product}\label{sec.LengthCircled}
Let us first assume $U=V$. As pointed out in Section \ref{sec.Jamiolkowski}, the Jamio\l kowski isomorphism $J$ induces on $\lvov$ an alternative product, which we denote with $\odot$. Using formula \eqref{fml.altprod.3}, one can easily prove that
\begin{equation}\label{circprodprod}
 \left(\left(A_1\otimes A_2\right)\odot\left(B_1\otimes B_2\right)\right)=\Tr\left(A_2^{T}B_1\right)\left(A_1\otimes B_2\right)
\end{equation}
for $A_i,B_j\in\lv$, $i,j=1,\ldots,d$. It is now straightforward to obtain the following
\begin{proposition}\label{prop.SepodotSep} Let $\rho$, $\sigma$ be separable states on $V\otimes V$ of lengths $l_1$, $l_2$, resp. Then $\rho\odot\sigma$ is proportional to a separable state of length $l\leqslant\min\left(l_1,l_2\right)$, with a proportionality coefficient $0\leqslant r\leqslant 1$.
\begin{proof}
 Let $\rho$ and $\sigma$ admit the following decompositions
\begin{equation}\label{decrhosigma}
 \rho=\sum_{i=1}^{l_1}p_i\rho^{\left(1\right)}_i\otimes\rho^{\left(2\right)}_i\quad\textrm{and}\quad\rho=\sum_{j=1}^{l_2}q_j\sigma^{\left(1\right)}_j\otimes\sigma^{\left(2\right)}_j
\end{equation}
for some $\rho^{\left(1\right)}_i, \rho^{\left(2\right)}_i, \sigma^{\left(1\right)}_j, \sigma^{\left(2\right)}_j\in\mv$ and positive numbers $p_i$, $q_j$ s.t. $\sum_{i=1}^{l_1}p_i=1$, $\sum_{j=1}^{l_2}q_j=1$. Assume w.l.o.g. that $l_1\leqslant l_2$, and thus $\min\left(l_1,l_2\right)=l_1$. From \eqref{circprodprod}, we get
\begin{equation}\label{circprodrhosigma}
 \rho\odot\sigma=\sum_{i=1}^{l_1}\sum_{j=1}^{l_2}p_iq_j\Tr\left(\rho_i^{\left(2\right)T}\sigma^{\left(1\right)}_j\right)\left(\rho^{\left(1\right)}_i\otimes\sigma^{\left(2\right)}_j\right):=\sum_{i=1}^{l_1}p_i\rho^{\left(1\right)}_i\otimes\sigma'^{\left(2\right)}_j,
\end{equation}
where we defined $\sigma'^{\left(2\right)}_j:=\sum_{j=1}^{l_2}q_j\Tr\left(\rho_i^{\left(2\right)T}\sigma^{\left(1\right)}_j\right)\sigma^{\left(2\right)}_j$. Since $0\leqslant\Tr\left(\rho_i^{\left(2\right)T}\sigma^{\left(1\right)}_j\right)\leqslant 1$, $\leqslant\sigma'^{\left(2\right)}_j$ is positive for all $j$ and $\Tr\sigma'^{\left(2\right)}_j\leqslant 1$. The proposition easily follows.
\end{proof}
\end{proposition}

\subsection{States of small lengths}\label{sec.smalllengths}
In this section we do not assume $U=V$ anymore. We will show that separable states of small lengths ($\leqslant 3$) necessarily have the length equal to their Schmidt rank. We also give an example of a separable state of length $4$ and Schmidt rank $3$.
\begin{proposition}\label{prop.smalllength}
 Let $\rho$ be a separable state on $U\otimes V$. Denote with $l$ the length of $\rho$ and with $r$ its Schmidt rank. If $l\leqslant 3$, $r=l$. 
For $l=1$, it is also true that $r=1\Leftrightarrow l=1$.
\begin{proof}
 If $l=1$, then $r$ must be equal to one. It turns out that the converse assertion also holds, that is $l=1$ if $r=1$ for a separable state $\rho$. Indeed, let $\rho=A_1\otimes B_1$ ($A_1\in\lu,B_1\in\lv$) be a separable. We know that $\rho$ is positive, so in particular $\hprodA{\rho}{u\otimes v}\geqslant 0\,\forall_{u\in U,v\in V}$. This is the same as $\hprodA{A_1}{u}\hprodA{B_1}{v}\geqslant 0\,\forall_{u\in U, v\in V}$, which implies that $A_1$ and $B_1$ are either both positive or both negative. If $A_1$, $B_1$ are positive, we obviously have $l=1$. If they are negative, we can write $\rho$ as $\left(-A_1\right)\otimes\left(-B_1\right)$, which is again a product of positive operators. So we conclude that $l=1$ whenever $r=1$ for a separable state $\rho$. It is actually not even necessary to assume separability of $\rho$. From the argument above it follows that a positive operator of Schmidt rank 1 is separable and of length 1. 



For $l=2$, the proof is almost immediate. Of course, we have $1\leqslant r\leqslant 2=l$. It is not possible that $r=1$, because this would imply $l=1$ (see the argument presented above). Therefore we conclude that $r=2$ whenever $l=2$.
 
We can proceed also for $l=3$, though this might be a little surprising at first. If $\rho$ has length 3, $\rho$ can be written as $\sigma_1\otimes\rho_1+\sigma_2\otimes\rho_2+\sigma_3\otimes\rho_3$, but it cannot be written as $\sigma_1\otimes\rho_1+\sigma_2\otimes\rho_2$ for some $\sigma_i$'s and $\rho_i$'s positive. We will show that the last two properties imply $r=3$. We know that $1\leqslant r\leqslant l=3$. If the sets \SEQ{\sigma_i}{i=1}{3} and \SEQ{\rho_i}{i=1}{3} are linearly independent, then obviously $r=3$, so we have to assume linear dependence of $\sigma_i$'s or $\rho_i$'s for $r$ to be less than 3.  Assume w.l.o.g. that the $\sigma_i$'s are linearly dependent. Thus $\sum_{i=1}^3\alpha_i\sigma_i=0$ for some sequence of numbers $\seq{\alpha_i}{i=1}{3}\subset\setR$ ($\sigma_i$'s are Hermitian, so their linear dependence over $\setC$ implies linear dependence over $\setR$). Because $\sigma_i$'s are all positive and different from zero, we must have at least one $\alpha_i$ negative and another $\alpha_i$ positive. We may assume w.l.o.g. $\alpha_1>0$ and $\alpha_2<0$. The sign of the remaining $\alpha_3$ can be arbitrary. We set $\alpha_3\leqslant 0$, but the proof for $\alpha_3\geqslant 0$ follows the same lines of argument. Denote $\beta_1:=\alpha_1$, $\beta_2:=-\alpha_2$ and $\beta_3:=-\alpha_3$. All the $\beta_i$'s are nonnegative and $\beta_1$, $\beta_2$ are strictly positive. The linear relation between $\sigma_i$'s reads
\begin{equation}\label{fml.lindep.1}
 \beta_1\sigma_1-\beta_2\sigma_2-\beta_3\sigma_3=0.
\end{equation}
Keeping in mind that $\beta_1\neq 0$, we can rewrite \fml{lindep.1} as
\begin{equation}\label{fml.lindep.2}
 \rho=\frac{\beta_2}{\beta_1}\sigma_2+\frac{\beta_3}{\beta_1}\sigma_3.
\end{equation}
Now \fml{lindep.2} implies that
\begin{equation}\label{fml.decofA.1}
 \rho=\sum_{i=1}^3\sigma_i\otimes\rho_i=\left(\frac{\beta_2}{\beta_1}\sigma_2+\frac{\beta_3}{\beta_1}\sigma_3\right)\otimes\rho_1+\sigma_2\otimes\rho_2+\sigma_3\otimes\rho_3,
\end{equation}
which is the same as
\begin{equation}\label{fml.decofA.2}
\rho=\sigma_2\otimes\left(\rho_2+\frac{\beta_2}{\beta_1}\rho_1\right)+\sigma_3\otimes\left(\rho_3+\frac{\beta_3}{\beta_1}\rho_1\right).
\end{equation}
Because of the nonnegativity of $\beta_i$'s 
and positivity of $\rho_i$'s, the operators $\rho_2+\frac{\beta_2}{\beta_1}\rho_1$ and $\rho_3+\frac{\beta_3}{\beta_1}\rho_1$ are positive, so from \fml{decofA.2} we have $l\leqslant 2$, which contradicts $r=3$. We conclude that $\sigma_i$'s cannot be linearly dependent. In a similar way, we can show that $\rho_i$'s cannot be linearly dependent. But linear independence of $\sigma_i$'s and $\rho_i$'s yields $r=3$, as we have already noticed above. We conclude that $r=3$ whenever $l=3$ for a separable state $\rho$. 
\end{proof}
\end{proposition}

The assertion of the above proposition is not true when $l>3$. We show this in the following example. 
\begin{example}\label{exp.l4r3}
 Consider $U=V=\setC^4$ and the following $4\times 4$ diagonal matrices
\begin{eqnarray}\label{defmatrE}
E_1=\diag{1,0,1,0},&E_2=\diag{0,1,0,1},\\
E_3=\diag{1,1,0,0},&E_4=\diag{0,0,1,1}.\nonumber
\end{eqnarray}
Let us identify $H_i$ with linear operators on $\setC^4$.
 Take $\rho\in\fnc{M}{\setC^4\otimes\setC^4}$ of the form
\begin{equation}\label{fml.exmplsep.2}
\rho:=\frac{1}{16}\sum_{i=1}^4E_i\otimes E_i. 
\end{equation}
The state $\rho$ is separable, has length 4 and Schmidt rank 3.
\begin{proof} 
 Obviously, $\rho$ is separable. For further convenience, let us denote the length of $\rho$ with $l$ and its Schmidt rank with $r$. We first prove that the Schmidt rank of $\rho$ is 3, which is equivalent to proving that the Schmidt rank of $\tilde\rho:=16\rho$ is 3. For that purpose we observe that the operators $E_i$ in \eqref{defmatrE} are linearly dependent. For example, we can write $E_4$ as a linear combination of $E_1$, $E_2$ and $E_3$,
\begin{equation}\label{fml.lindepEi.1}
 E_4=E_1+E_2-E_3.
\end{equation}
We can put \fml{lindepEi.1} in \fml{exmplsep.2} and use distributivity of $\otimes$ to get
\begin{equation}\label{fml.exmplsep.4}
\tilde\rho=E_1\otimes\left(2E_1+E_2-E_3\right)+E_2\otimes\left(E_1+2E_2-E_3\right)+E_3\otimes\left(2E_3-E_1-E_2\right),
\end{equation}
From \fml{exmplsep.4}, we definitely see that $\tilde\rho$ has Schmidt rank lower than four. But the matrices $2E_1+E_2-E_3$, $E_1+2E_2-E_3$ and $E_1+E_2-2E_3$ are linearly independent\footnote{the matrix $\left[\begin{array}{ccc}2&1&-1\\1&2&-1\\1&1&-2 \end{array}\right]$ has a nonzero determinant}, just as the matrices $E_1$, $E_2$ and $E_3$ are. This implies that the number of product terms in \fml{exmplsep.4} cannot be reduced any further. Consequently, the Schmidt rank of $\tilde\rho$ and of $\rho$ is 3, $r=3$.

Of course, the length of $\rho$ is not lower than $r$, so we have $l\geqslant 3$. On the other hand, \fml{exmplsep.2} is an expression for $\rho$ as a sum of four products of positive operators $E_i$. Therefore $l$ cannot be higher than 4 and the only possibilities left are $l=3$ and $l=4$. In the following we show that $l=3$ is excluded. Put it in a different way, $\rho$ cannot be written as
\begin{equation}\label{fml.decofC2.1}
 F_1\otimes G_1+F_2\otimes G_2+F_3\otimes G_3
\end{equation}
with $F_i$ and $G_i$ positive for $i=1,2,3$. It will be more convenient to show that $\tilde\rho$ cannot be written in the form \eqref{fml.decofC2.1} with all $G_i$, $F_i$ positive. To prove this, let us assume that a decomposition of the form \fml{decofC2.1} exists. We should stress that \fml{exmplsep.4} is not an example of such a decomposition because $2E_3-E_1-E_2$ is not positive. The operators $F_i$ and $G_i$ are elements of $\fnc{H}{\setC^4}$, so we can write them as $F_i=\sum_{j=1}^{16}\alpha_i^jH_j$ and $G_i=\sum_{j=1}^{16}\beta_i^jH_j$, where $\alpha_i^j,\beta_i^j\in\setR\,\forall_{i,j}$, \SEQ{H_j}{j=1}{16} is a basis of $\fnc{H}{\setC^4}$ such that
\begin{eqnarray}\label{fml.defmatrH.1}
 H_1=\diag{1,0,0,0},&H_2=\diag{0,1,0,0},\\
H_3=\diag{0,0,1,0},&H_4=\diag{0,0,0,1}.\nonumber
\end{eqnarray}
and $H_j$'s have only off-diagonal nonzero elements for $j\geqslant 5$. Because of the form \eqref{defmatrE} of the operators $E_i$, $\tilde\rho$ does not have any off-diagonal elements and the decomposition of $\tilde\rho$ in the basis \SEQ{H_k\otimes H_l}{k,l=1}{16} of $\fnc{H}{\setC^2\otimes\setC^2}=\fnc{H}{\setC^2}\otimes\fnc{H}{\setC^2}$ does not include any terms with $k\geqslant 5$ nor with $l\geqslant 5$. If there are any terms including $H_k$ with $k\geqslant 5$ in $F_i$ or $G_i$, they must eventually cancel out in the tensor product \fml{decofC2.1}. Therefore we may use $\widetilde F_i:=\sum_{j=1}^4\alpha_i^jH_j$ and $\widetilde G_i:=\sum_{j=1}^4\beta_i^jH_j$ instead of $F_i$ and $G_i$. The relation \fml{decofC2.1} still holds when $F_i$ is replaced with $\widetilde F_i$ and $G_i$ with $\widetilde G_i$. Positivity of $\widetilde F_i$ and $\widetilde G_i$ follows from the fact that they are diagonal parts of positive operators. We see that $\sum_{i=1}^4\widetilde F_i\otimes\widetilde G_i$ equals $\tilde\rho$, but it is also a sum of products of positive operators. Consequently, if there exists a decomposition of $\tilde\rho$ of the form \fml{decofC2.1} with $F_i$ and $G_i$ positive, another decomposition with diagonal and positive $F_i$ and $G_i$ must also exist. Therefore we can just concentrate on decompositions of the form
\begin{equation}\label{fml.decofC2.2}
 \tilde\rho=\sum_{i=1}^3\sum_{j,k=1}^4\alpha_i^jH_j\otimes\beta_i^kH_k=\sum_{i=1}^3\sum_{j,k=1}^4\alpha_i^j\beta_j^kH_j\otimes H_k
\end{equation}
with $\alpha_j^j\geqslant 0$ and $\beta_i^k\geqslant 0$. It can be easily checked that
\begin{equation}\label{fml.decofC2.3}
 \tilde\rho=\sum_{j,k=1}^4A^{jk}H_j\otimes H_k
\end{equation}
with $A^{11}=A^{22}=A^{33}=A^{44}=2$, $A^{14}=A^{41}=A^{23}=A^{32}=0$ and $A^{ij}=1$ for the remaining eight $A$ coefficients. In order for \fml{decofC2.2} to hold, we must have
\begin{equation}\label{fml.coeffsofC2.1}
\sum_{i=1}^4\alpha_i^j\beta_i^k=A^{jk}\,\forall_{j,k\in\left\{1,2,3,4\right\}}.
\end{equation}
To see the consequences of \fml{coeffsofC2.1}, let us introduce vectors $\alpha^j\in\setR^3$ and $\beta^k\in\setR^3$ with coordinates \SEQ{\alpha_i^j}{i=1}{3} and \SEQ{\beta_i^k}{i=1}{3}, respectively. The conditions \fml{coeffsofC2.1} can be written as
\begin{eqnarray}
 \alpha^1\cdot\beta^1=\alpha^2\cdot\beta^2=\alpha^3\cdot\beta^3=\alpha^4\cdot\beta^4=2\label{fml.alphabetaprodC2.1},\\
\alpha^1\cdot\beta^4=\alpha^4\cdot\beta^1=\alpha^2\cdot\beta^3=\alpha^3\cdot\beta^2=0\label{fml.alphabetaprodC2.2},\\
\alpha^1\cdot\beta^2=\alpha^1\cdot\beta^3=\alpha^4\cdot\beta^2=\alpha^4\cdot\beta^3=1\label{fml.alphabetaprodC2.3},\\
\alpha^2\cdot\beta^1=\alpha^2\cdot\beta^4=\alpha^3\cdot\beta^1=\alpha^3\cdot\beta^4=1\label{fml.alphabetaprodC2.4}.
\end{eqnarray}
Keeping in mind nonnegativity of $\alpha_i^j$'s and $\beta_i^k$'s, we can draw some further conclusions about these numbers. First of all, we should notice that two real vectors with nonnegative coordinates are orthogonal if and only if a nonvanishing coordinate of one of the vectors corresponds to a vanishing coordinate of the other vector and vice versa.  As a consequence of this and \fml{alphabetaprodC2.2}, each of the vectors $\alpha^i$ and $\beta^i$ must have a vanishing coordinate. On the other hand, because of the formula \fml{alphabetaprodC2.1} neither of the vectors can be zero. In other words, each of them must have a nonvanishing coordinate. We are left with $\alpha^i$'s and $\beta^j$'s which have either one or two nonzero coordinates. Let us consider first a situation in which one of the vectors has two nonzero coordinates. Without any loss of generality we assume the vector to be $\alpha^1$ and we put $\alpha_1^1=0$, $\alpha^1_2>0$, $\alpha^1_3>0$. Because of \fml{alphabetaprodC2.2}, $\beta_1^4>0$, $\beta_2^4=0$, $\beta_3^4=0$. This in turn implies $\alpha_1^2>0$, $\alpha_1^3>0$ and $\alpha_1^4>0$ as a consequence of \fml{alphabetaprodC2.1}, \fml{alphabetaprodC2.3} and \fml{alphabetaprodC2.4}. Therefore $\beta_1^3=0$, $\beta_1^2=0$ and $\beta_1^1=0$. If $\alpha_2^2=\alpha_3^2=0$, the equality $\alpha^2\cdot\beta^1=1$ cannot hold. One of the coordinates $\alpha_2^2$, $\alpha_3^2$ must be nonzero. We may assume $\alpha_3^2>0$, so that we have $\alpha_1^2>0$, $\alpha_2^2=0$, $\alpha_3^2>0$. From \fml{alphabetaprodC2.2} it follows that $\beta_1^3=0$, $\beta_2^3>0$, $\beta_3^3=0$. Using \fml{alphabetaprodC2.1} we get $\alpha_2^3>0$ while \fml{alphabetaprodC2.3} yields $\alpha_2^4>0$. We have obtained $\alpha_1^3>0$ and $\alpha_2^3>0$, which implies $\alpha_3^3=0$. But now \fml{alphabetaprodC2.2} gives us $\beta_1^2=0$, $\beta_2^2=0$, $\beta_3^2>0$ and from $\alpha^4\cdot\beta^2=1$ we get $\alpha_3^4>0$.

In the successive steps above we obtained $\alpha_1^4>0$, $\alpha_2^4>0$ and finally $\alpha_3^4>0$. This is in contradiction with \fml{alphabetaprodC2.2}, so our initial assumption about the existence of a vector $\alpha^i$ (or $\beta^i$) with two nonzero coordinates, cannot be true for solutions of the equations \fml{alphabetaprodC2.1}-\fml{alphabetaprodC2.4}. None of the vectors $\alpha^i$, $\beta^i$ can have two nonvanishing coordinates. The only possibility we have not excluded yet is that of all the vectors $\alpha^i$, $\beta^i$ having precisely one nonzero coordinate each. Let us assume that this is the case and concentrate on $\alpha^i$'s. Because of the fact that $\alpha^i$'s are of dimension three, there must be a pair of indices $i\neq j$ such that $\alpha^i=\alpha^j$. Without any loss of generality we may assume that either $\alpha^1=\alpha^2$ or $\alpha^1=\alpha^4$ holds. The first possibility is excluded because of the equalities $\alpha^1\cdot\beta^4=0$ and $\alpha^2\cdot\beta^4=1$.  The second is in contradiction with $\alpha^1\cdot\beta^4=0$ and $\alpha^1\cdot\alpha^1=2$. Thus we have excluded the only remaining possibility for $\alpha^i$'s and we conclude that \fml{coeffsofC2.1} has no solutions of the desired properties $\alpha_i^j,\beta_i^k\geqslant 0\,\forall_{i,j,k}$. Consequently, $\tilde\rho$ cannot be written in the form \fml{decofC2.1} with $F_i$'s and $G_i$'s positive. The same holds for $\rho$. Hence $l>3$, which in turn implies $l=4$ because $l\leqslant 4$. This proves our assertions about $\rho$.
\end{proof}
\end{example}

\section{Conclusion}

In Section \ref{sec.ConeDual}, we used duality relations and other specific properties of the cones of $k$-positive and $k$-superpositive maps to obtain analogues of the positive maps criterion by Horodeccy \cite{ref.Horodeccy}.
Our results can in particular be applied to $2$-positive maps, which are very closely related to the set of undistillable states, \cite{Cl05}. As we explained in Section \ref{sec.BPPospolys}, block positivity of operators is equivalent to a system of inequalities involving real homogenous polynomials. We showed an example (cf. formula \eqref{abcfamilydef}) where these can be solved explicitly, thus yielding explicit conditions for block positivity. We also proved in Section \ref{sec.neccesaryimsufficient} that certain type of approach to the block positivity question can never result in a sufficient criterion. We also touched upon the problem of block positivity over $\setR$ and its relation to sums of squares (SOS). Finally, in Section \ref{sec.cardinality} we introduced the notion of length of a separable state and discussed its relation to the Schmidt rank for states of small lengths. Length may be considered as a measure of the amount of classical communication needed for the creation of a separable quantum state.

Many questions that arised during the preparation of this paper remain open. Let us mention a few of them:
\begin{itemize}
 \item Does the family of symmetric mapping cones contain all the cones that satisfy an analogue of the positive maps criterion? 
\item What is the exact relation between SOS and entanglement witnesses, e.g. to their decomposability? 
 \item Does there exist a separable state of length 4 with Schmidt rank two?
 \item Can we fully characterize separable states of small lengths?
\end{itemize}

\section*{Acknowledgement}
The author would like to thank Karol \.Zyczkowski for his support during the preparation of the Master's Thesis and for suggesting new research topics.  Erling St\o rmer has greatly contributed to the preparation of \cite{SSZ09}, which served as a basis for Section \ref{sec.ConeDual}. Very helpful comments written by Ronan Quarez allowed the author to better understand questions related to polynomial positivity, touched upon in Section \ref{sec.BPoverreals}. Discussions with Pawe\l\ Horodecki are kindly acknowledged. Finally, the author is sincerely thankful for the invitation by Berthold-Georg Englert to submit the thesis to IJQI. The project was initiated in Singapore, where the author enjoyed hospitality at the National Technical University during the Les Houches School of Physics in Singapore (Session XCI). The paper was brought into its final form during a visit of the author to the University of Oslo, when he was financially supported by Scholarship and Training Fund, operated by Foundation for the Development of the Education System. Hospitality of the Mathematics Institute of the University of Oslo is also gratefully acknowledged.
\vskip 3 mm
Project operated within the Foundation for Polish Science International Ph.D. Projects Programme co-financed by the European Regional Development Fund covering, under the agreement
no. MPD/2009/6, the Jagiellonian University International Ph.D. Studies in
Physics of Complex Systems.

\end{document}